\documentclass[amssymb, amsmath, aps, pre]{revtex4-2}

\usepackage{amsmath}%
\usepackage{amssymb}
\usepackage{dcolumn}
\usepackage{bm}
\usepackage{color}
\usepackage{mathrsfs}
\usepackage{epsfig}
\usepackage{bm}

\begin{document}

\title{Cascade models of anisotropic turbulence in magnetized solar wind and astrophysical plasmas}

\author{A. Bershadskii}

\affiliation{
ICAR, P.O. Box 31155, Jerusalem 91000, Israel
}

\begin{abstract}

We model the solar wind plasma as a collection of discrete domains, each possessing local rotational symmetry about its domain-mean magnetic field. Based on this partial symmetry, we construct cascade models governed by the nonlinear and Alfvénic time scales. We introduce a stochastic renewal process with a bimodal waiting-time distribution determined by these two time scales. This process yields a two-branch renormalization solution for the total energy: a statistically robust branch with an Iroshnikov-Kraichnan-like  $k^{-3/2} $ spectrum and a statistically marginal branch with a Kolmogorov-like  $k^{-5/3} $ spectrum. Whenever energy-flux fluctuations become supercritical, the system selects the faster of the two available cascade rates, thereby preventing flux accumulation. During solar wind expansion, the system undergoes a dynamical first-order phase transition from the slow  $k^{-3/2} $ cascade to the fast  $k^{-5/3} $ cascade, with switchbacks accelerating the transition. The mixing of subcritical and supercritical domains in raw spacecraft measurements produces an apparent decoupling of the velocity and magnetic-field spectral indices. Incorporating a magnetic helicity cascade slaved to the energy cascade, we show that the large-scale spectra decouple into  $k^{-3/4} $ magnetic and  $k^{-3/2} $ kinetic spectra. For switchbacks inside the upper solar corona, the corresponding large-scale spectra instead decouple into  $k^{-2/3} $ magnetic and  $k^{-5/3} $ kinetic spectra. Data from Voyager, Ulysses, Helios, Wind, and Parker Solar Probe confirm these spectral signatures across heliospheric regions spanning heliocentric distances from 0.056 to 140 AU. Finally, we apply this framework to the interstellar and intracluster media to show universality of the cascade models.

\end{abstract}

\maketitle

\section{Introduction}

   Standard turbulence theory was built on symmetry. Kolmogorov's theory of hydrodynamic turbulence, and its first magnetohydrodynamic (MHD) descendants, rest on assumptions of rotational symmetry (isotropy), translational symmetry (homogeneity), directional balance, and invariance under scale transformations (scaling). These assumptions are not incidental simplifications; they are what makes the theory tractable at all. Yet the solar wind -- the one turbulent astrophysical plasma we can sample directly, in situ, for decades on end -- violates every one of them in some measure. This paper examines each symmetry in turn, asks what survives its violation, and argues that what survives is real: on a deeper statistical level the symmetries persist, but only once the naive, global form of each is relaxed and replaced by something more adapted to the physics of an expanding, magnetized, patchy medium.\\

The clearest case is isotropy (rotational symmetry). Standard MHD turbulence theories struggle to explain the persistent, concurrent appearance of Kolmogorov-like ($k^{-5/3}$) and Iroshnikov--Kraichnan (IK)-like ($k^{-3/2}$) spectra in the solar wind -- the two central results of twentieth-century turbulence theory, both originally derived for isotropic and homogeneous turbulence. When the $k^{-5/3}$ spectrum turned up in demonstrably anisotropic hydrodynamic flows, Kolmogorov rescued the theory with the idea of local isotropy \cite{my}: turbulence need not be globally isotropic if, at each point, it becomes isotropic when viewed relative to its own local reference frame. This move works even in ordinary shear flows, which do have a globally preferred direction, because the underlying eddies are free to reorient: the "fluidity" of vortex turbulence allows progressively more complete randomization of orientation as scale decreases, so that memory of the large-scale direction is gradually lost toward smaller scales. It works far less well in the solar wind, because a strong mean (background) magnetic field does not simply bias vortices toward some direction -- it replaces them, as the dominant small-scale structure, with Alfvén wave packets that are intrinsically tied to the field direction and have no comparable freedom to reorient. There is no local frame, at any scale, in which that constraint relaxes. The growing weight of contradictions, in both numerical simulations and in situ observations, has gradually forced the community to treat Alfvénic solar-wind turbulence as three-dimensionally anisotropic in both the global and the local sense. And yet both spectra are persistently observed. The current phenomenological accounts split along these lines: the $k^{-5/3}$ spectrum is generally attributed to a "Critical Balance" cascade \cite{gs,lgs}, the $k^{-3/2}$ spectrum to a "Dynamical Alignment" cascade \cite{bold}. Both are contested, and their application to the solar wind specifically has been strongly criticized; the resulting puzzle was reviewed comprehensively in \cite{sch}. Part of the resolution, as the third section below argues, may not lie in choosing between these two cascade pictures at all, but in recognizing that the wind itself is not a single homogeneous population sampling one cascade -- a point that reframes the isotropy problem as, at bottom, a homogeneity problem. Also, the isotropy collapses from a genuinely three-dimensional to an effectively two-dimensional one restricted to the plane transverse to the mean magnetic field $\bf{B}_0$ (the highly debated issue of tubes vs ribbons \cite{sch}).\\

  A second, related symmetry concerns direction along the mean field rather than around it: the Alfvénic imbalance problem. This is a consequence of the solar wind's inherent anisotropy, and it too is far from solved \cite{sch,lgs,chadran,bl,pb}. In a state of imbalanced turbulence, magnetic and velocity fluctuations propagating in one sense (predominantly outward from the Sun) carry substantially more energy than those propagating in the other (inward), which breaks the symmetry that standard counter-propagating-wave pictures of the cascade assume. The imbalance is often most pronounced in young solar wind close to the Sun. It has recently been suggested \cite{yang} that strongly imbalanced turbulence is sustained primarily by coherent interactions between the dominant outward waves and fluctuations that co-propagate with them, rather than by the counter-propagating wave collisions that standard cascade phenomenology assumes. As the wind expands, nonlinear wave reflection gradually drives the turbulence toward better balance, but full balance is never reached, even far from the Sun. Here again, though, the interesting question is not whether perfect balance holds -- it does not -- but whether the degree and radial evolution of imbalance settle onto some more universal statistical relationship, a possibility taken up later in this paper.\\
  
  A third symmetry problem concerns translational invariance, or homogeneity, and is in some ways the deepest of the four, because it bears directly on how the other three are even measured. The solar wind is fragmentary, or patchy: different patches carry different physical properties. Patchiness is an inherent feature of chaotic, turbulent motion generally, and it is especially prominent in MHD turbulence (see, e.g., \cite{b2,b4}). In the solar wind, it has an additional, external source: different regions of the Sun, at different times, inject plasma of different character into the wind, layering solar-surface structure on top of the turbulence's own intrinsic patchiness. This compounds the difficulty of interpreting spacecraft data, which arrive as time series and are converted to spatial structure via Taylor's frozen-in-flow hypothesis -- treating successive intervals as though they were spatial patches advected past the probe at the bulk solar-wind speed. Under this interpretation, an average over a time series is implicitly an average over patches that may have genuinely different physical properties, and that average need not represent any of them. This is not merely a statistical nuisance; it plausibly bears directly on the first symmetry problem above. If some patches favor conditions consistent with a $k^{-3/2}$ spectrum and others favor $k^{-5/3}$, the spectrum recovered from an interval spanning both will show one, the other, or something in between, depending on the relative weight of each population in the sample -- a mixing effect that could account for part of the persistent, unresolved coexistence of the two spectral laws discussed above, independent of any question about which cascade theory is "correct." Several sampling techniques have been proposed to disentangle patches from genuine turbulent structure (see, e.g., \cite{hor,wick1,wick2}), and these are discussed later in this paper.\\
  
  A fourth symmetry is more sophisticated.  In the solar wind, a mean (background) magnetic field $\bf{B}_0$ breaks the relabeling symmetry responsible for magnetic helicity $H=\int\mathbf{A}\cdot\mathbf{B}\,dV$ (where ${\bf B} = \nabla \times {\bf A}$) conservation. This symmetry consists of local, field-line-wise permutations of fluid-element labels constrained only to be constant along $\bf{B}$ \cite{pm}; it is the Noether source of $H$-conservation, distinct from the gauge invariance of $\bf{A}$, which merely ensures $H$ is well-defined. When $\bf{B}_0$ dominates, this local freedom collapses from a genuinely three-dimensional, field-line-resolved symmetry to an effectively two-dimensional one restricted to the plane transverse to $\bf{B}_0$ (as for the above-considered rotational symmetry). Total helicity remains formally conserved, but the symmetry that protects it -- and hence its role in constraining the turbulence -- is no longer the full one available at $\bf{B}_0=0$ \cite{shebalin}.

   The breaking is not total, and what survives is what makes a cascade of fluctuating helicity possible. Writing $\bf{B}=\mathbf{B}_0+\mathbf{b}$, $\bf{A}=\bf{A}_0+\bf{a}$, the total helicity splits into a fluctuation term $H_f=\int {\bf a}\cdot{\bf b}\,dV$, a cross term $2\int{\bf A}_0\cdot {\bf b}\,dV$, and a mean-field self-term. Because ${\bf A}_0$ is not periodic, the cross term is gauge-origin dependent and precludes a well-defined local decomposition of the total $H$ \citep{bf,fa}. The fluctuation helicity $H_f$, however, retains a residual gauge freedom ${\bf a}\to {\bf a}+\nabla\chi$ under periodic $\chi$, leaving $H_f$ exactly gauge-invariant regardless of ${\bf B}_0$. This residual invariance is what permits the scale-to-scale cascade analysis of magnetic helicity fluctuations developed below. \\

     A fifth symmetry, invariance under scale transformations, or scaling, has been the most productive tool in turbulence studies since Kolmogorov, precisely because power-law spectra are the observable signature of scale invariance. Recent observations, however, show this symmetry a serious competitor in the solar wind: stretched-exponential spectra, associated with what is termed distributed chaos \cite{b1}, in which the spectral energy falls off faster than any power law, and the departure from scale invariance is itself governed by an underlying statistical distribution of local dissipation rates, rather than by a single dominant scale. Scaling symmetry is not simply absent where distributed chaos appears; it persists, but one level further down, in the statistics that govern the departure of the spectra from a pure power law. The present paper will demonstrate that the two approaches are closely connected.\\
     
   Taken together, these five cases point to a common answer to the underlying question of whether classical symmetry arguments have any purchase on the real solar wind. They do -- but not in the strong, global form in which they were first formulated. On a deeper statistical level, rotational symmetry, directional balance, homogeneity, rescaling symmetry, and scale invariance are all still present in some relaxed, often local or statistical sense. But the real solar wind is considerably more complex than the elegant restrictions these symmetries impose in their idealized form, and progress in the field has consistently come from relaxing those restrictions and adapting them with other, complementary physical methods.\\
   
  In this paper, we turn this statistical perspective into a quantitative cascade framework. Modeling the turbulent medium as an ensemble of discrete domains (each retaining local rotational symmetry about its domain-mean magnetic field), we formulate a stochastic renewal process governed by a bimodal waiting-time distribution built from the nonlinear and Alfvénic time scales. We show that this process yields two renormalization branches connected by a dynamical first-order phase transition triggered when energy-flux fluctuations become supercritical, and we test the resulting predictions both qualitatively and quantitatively against multi-spacecraft observations spanning heliocentric distances from 0.056 to 140 AU (Parker Solar Probe, Helios, Wind, Ulysses, and Voyager). \\
 
  Finally, incorporating a magnetic helicity cascade slaved to the energy cascade resolves the dynamics of the large-scale solar wind -- the energy-containing reservoir. Beyond the heliosphere, this same framework extends naturally to galactic and intracluster plasmas, providing a universal MHD description for magnetized plasmas that differ vastly in scale and physical properties.

\section{Minimalistic cascade models}

\subsection{Minimalistic approach for the Kolmogorov cascade}

 Let us begin with a minimalistic model of the Kolmogorov cascade in isotropic and homogeneous turbulence. The solar wind violates both symmetries: isotropy (rotational symmetry) and homogeneity (translational symmetry). However, the core of this model and the ideas we will use to justify it will be instrumental in further considerations of the solar wind.
 
  Let us define a time interval corresponding to a single energy cascade step $\tau_{cas}= \ell/v^2$, where $  v^2$ is the kinetic energy fluctuation (over the scale of reference $\ell$) which will be transferred over a single cascade step. This definition is based on the 3D isotropy (rotational symmetry). It will be appropriate for the estimation of the constant energy flux $\varepsilon \sim   v^2 /\tau_{cas}$. Let us consider the constant mean energy flux $\varepsilon$. Then,
\begin{equation}
\varepsilon = \frac{v^2}{\tau_{cas}} = \frac{v^2}{\ell / v} = \frac{v^3}{\ell} \implies v^2 \sim (\varepsilon \ell)^{2/3}.
\end{equation}
  Because the spectral energy density $E(k)$ represents energy per unit wavenumber ($k \sim 1/\ell$), the total energy scale is related to the spectrum via $  v^2 \sim k E(k)$. Therefore,
\begin{equation}
k E(k) \sim \varepsilon^{2/3} k^{-2/3} \implies E(k) \propto \varepsilon^{2/3} k^{-5/3}.
\end{equation}

This model has two main points of vulnerability: the specific choice of the $\tau_{cas}= \ell/  v$ and the independence of $\varepsilon$ on $\ell$. 
   Immediately after its appearance in 1941, Kolmogorov's phenomenological cascade approach was criticized by L. D. Landau. The criticism was based on a very reasonable statement that, in real turbulent flows, the energy flux must fluctuate. This criticism forced A.N. Kolmogorov to create an intermittency theory in 1961, which was also strongly criticized. In hydrodynamics, this problem is still very far from a solution. Paradoxically, this phenomenon is really not a problem but a key to the justification of this approach. Indeed, in an incompressible fluid there is a causality limitation on $\tau_{cas} \geq \ell/ v $, because in this fluid energy and momentum cannot propagate with a speed larger than $v$. The spatio-temporal fluctuations of the local energy flux  $\varepsilon  $ lead to the intermittent accumulation of energy at specific scales  $\ell  $. This local concentration of energy can cause cascade instabilities unless the energy is sufficiently rapidly transferred to smaller scales via inertial vortex stretching. The lower causal limit $\tau_{cas} = \ell/ v$ corresponds to the fastest cascade. Therefore, it should be chosen for $\tau_{cas}$. It will be shown below that the two self-supporting fundamental principles suggested here -- causality and stability can be successfully applied for an analysis of the highly anisotropic and inhomogeneous MHD turbulence in the solar wind. 

\subsection{Causal time scales and restrictions on the minimalistic MHD models}

  The minimalistic model can be generalized to account for certain significant realities of the Alfvénic solar wind under conditions of deterministic causality. The main nonlinear interactions in the Alfv\'enic solar wind are supposed to be between the Alfv\'en wave packets (AWPs) propagating along the background magnetic field ${\bf B}_0$. First paradox of the Alfv\'enic theoretical description is that the usually used ``nonlinear'' characteristic time scale $\tau_{nl} = \tau_{\perp} = \ell_{\perp}/v_{\ell}$ (where $\ell_{\perp}$ is the scale of reference in direction perpandicular to ${\bf B}_0$) and does not account for the contribution of the magnetic nonlinear term. This fact can be attributed to the idea that just the nonlinearity of the velocity term in the momentum equation is the main source of chaotic (turbulent) processes in MHD. This is also supported by the straightforward causal meaning of such defined $\tau_{nl}$ (which cannot be made using the fluctuations of the magnetic field ${\bf b}$). Although,  a characteristic (causally meaningful) time scale based on the linear part of the Lorentz force  $\tau_{\parallel} = \tau_A = \ell_{\parallel}/V_A$ (where $\bf{V}_A = \bf{B}_0/\sqrt{\mu_0 \rho}$ is the Alfv\'en velocity, $\mu_0$ is the permeability of free space, and $\rho$ is the plasma density) can be well defined, because the $V_A$ is the velocity of the Alfv\'enic  wave packets (AWPs) propagation along the bacground magnetic field ${\bf B}_0$.\\
  
  The definition of the $\ell_{\perp}$ and $\ell_{\parallel}$ also needs certain clarification. The solar wind is fragmentary (inhomogeneous). Two main sources of patchiness are the origin of the solar wind from the rather variable in space and time solar atmosphere, and the inherent spontaneous patchiness of MHD turbulence. The direction of the local mean (over the patch) magnetic field ${\bf B}$ is different in different spatial patches (domains). The primary anisotropy, relevant to the nonlinear interactions between the contrapropagating along ${\bf B}$ Alfv\'en wave packets (AWP), is usually described by the angle between the bulk velocity of the solar wind ${\bf V}$ (which has predominantly radial direction) and {\bf B}: $\theta_{VB}$. \\
  
   Since the main turbulence activity in the magneto-inertial range of scales in the Alfv\'enic solar wind is supposed to be provided by interaction (colliding) of the contrapropagating along the mean magnetic field Alfv\'enic wave packets, we will use the effective cross-section to describe these collisions. The wave-particle (quasi-particle) duality is a well-known phenomenon in quantum physics. 

Then, $\ell_{\parallel} = \ell |\cos \theta_{VB}|$ and $\ell_{\perp} =\ell |\sin \theta_{VB}|$ (where $\ell \ll L$ is the scale of reference along ${\bf V}$, and $L$ is the characteristic scale of the patch). The measurements by a probe (aboard a spacecraft) are usually obtained as a time series. Using Taylor's hypothesis, corresponding frequency spectra $E(f)$ of the energy fluctuations are usually interpreted as 1D wavenumber spectra $ E (k) $ with the replacement  $k \propto f/V$. Choosing $k \propto 1/\ell$ and the relation $E(k)k \sim   v^2$ we can find the 1D spectrum $E(k)$ if we know $ v^2$.\\

  To avoid the common conflation between the fluctuations belonging to the magneto-inertial range of scales and local mean magnetic fields, we define a strict two-scale hierarchy. The vector $\mathbf{B}$ represents the local mean magnetic field of a macroscopic spatial domain (patch scale $L \gg \ell$), which dictates the parallel Alfv\'enic transit time $\tau _{\parallel}$ through $V_A$ and the angle $\theta_{VB}$. Conversely, $v_{\ell}^2$  denotes the stochastically fluctuating kinetic energy of a singular fluctuation at the magneto-inertial scale of reference $\ell$. While local configurations of $v_{\ell}^2$  can induce localized ribbon-like structures within an isolated wave packet (the well-known controversy ``tubes vs ribbons'' \cite{sch}), the localized geometrical and kinematic anisotropies have a second-order (microscopic) effect on the energy cascade in the patch of size $L \gg \ell$ in comparison to the first-order anisotropic effect described by the macroscopic angle $\theta_{VB}$. Therefore, the effective perpendicular time $\tau_{\perp}$ can be considered approximately isotropic in the plane perpendicular to ${\bf B}$ in the cascade's first-order approximation.\\
  
  Crucially, the effective cross-section approach mirrors the exact conditional sampling methodologies employed in space plasma observations. When in-situ measurements are binned by the angle $\theta _{VB}$ to extract directional spectra, the resulting curve is not a snapshot of a single, isolated domain. Instead, it represents an explicit ensemble average over thousands of independent spatial patches that happen to share the same macroscopic orientation of $\mathbf{B}$. This provides an ensemble average, additionally justifying the use of the effective cross-section. \\
  
  The introduced definitions and assumptions are not sufficient to obtain a closed minimalistic model of an MHD energy cascade. Generally, there is an active exchange between kinetic and magnetic energies, while the total energy (kinetic plus magnetic) is conserved in the ideal (non-dissipative) MHD. Therefore, it is the total energy that should cascade. Here, the above-mentioned paradox with $\tau_{nl}$ determined as $\tau_{nl} = \ell_{\perp}/v_{\ell}$ plays a crucial role, apparently contradicting the subject of cascade. To solve this problem, it was suggested in the seminal paper \cite{kr} -- the Iroshnikov-Kraichnan (IK) approach -- that there is an equipartition between kinetic and magnetic fluctuations at every scale in the inertial range $v_{\ell} \approx b_{\ell}$ (in the Alfv\'enic units). This hypothesis is still popular in the MHD literature, although the numerical simulations of the Alfv\'enic turbulence and measurements in the Alfv\'enic solar wind do not support it.  
  
  This assumption is usually motivated by analogy -- a shear linear Alfv\'en wave has exactly equal kinetic and magnetic energy -- but an analogy to the linear eigenmode does not by itself tell us how rigid this constraint is inside the nonlinear cascade argument, or how much it could be loosened without breaking IK's own logic. That is the question we set out to answer: not "does equipartition hold," but what is the widest closure $v_{\ell} = C(\ell)b_{\ell}$ that the IK derivation itself can tolerate, without silently importing new physics? This matters because the answer determines whether departures from equipartition -- such as the observed magnetic excess (residual energy) in solar-wind inertial range turbulence -- can be accommodated as a refinement of IK, or whether they necessarily require a mechanism external to it.

  MHD turbulence in the strong-mean-field regime has exactly two causal time-scales available: the eddy turnover $\tau_{nl} = \ell_{\perp}/v_{\ell}$ and the Alfvén time $\tau_A = \ell_{\parallel}/V_A$. The induction equation shows that ${\bf b}$ is sourced by ${\bf v}$ through $({\bf B}_0 \cdot \nabla) {\bf v}$, acting coherently for one $\tau_A$. Over a single coherence step, this forcing gives $b_{\ell} \approx   v_{\ell}/C_{\ell}$, with $C_{\ell}$ an $O(1)$ number fixed by local geometry — not pinned to any specific value by the argument. The $C_{\ell}$ can `adiabatically'' change. ``Adiabatic'' here carries its usual physical meaning: $C(\ell)$ is allowed to change across the inertial range, but only slowly (compared to the local cascade dynamics at scale $\ell$ itself). Over one cascade step, $C$ changes by a small, cumulative fractional amount, not a considerable finite jump. This is the same sense of "adiabatic" as in a slowly-varying WKB envelope: at any given scale, the local physics still looks exactly like IK's single-step balance with an effectively constant $C$, and the drift only becomes visible after accumulating over many steps. If $C$ instead drifted fast -- a considerable finite fractional change per step -- that would be equivalent to smuggling in a third causal time, i.e., new physics outside the IK approach (e.g., breaking the isotropy in the perpendicular plane -- a ribbon-like structure). Adiabatic drift is precisely the boundary case that stays inside IK's two-causal-times structure while still being nontrivial.

  A slow, cumulative drift built from many small, independent per-step changes is a multiplicative random walk -- the same structure underlying Kolmogorov's refined similarity hypothesis (intermittency) \cite{my}, where the local dissipation rate $\varepsilon_{\ell}$ drifts adiabatically rather than being reset by new physics at each scale. This can be a source of the magnetic field intermittency. We will consider how far we can go with this hypothesis in the first part of the paper (without exploring the intermittency effect) and then change it to account for a strong influence of another ideal MHD invariant -- magnetic helicity.
  
\subsection{Deterministic causal model}

     In the solar wind, we have two causal windows $\tau_{\perp}$ and $\tau_{\parallel} $ . Therefore $\tau_{cas} \geq \max\{\tau_{\perp}, \tau_{\parallel}\}$. Usually in the magneto-inertial range the turbulence is weak: the the total energy fluctuation $ \mathcal{E}_{\ell} \sim (v^2 +b^2) \sim v^2 \ll V_A^2 \implies \tau_{\perp} \gg \tau_{\parallel}$ except of a tiny vicinity of the angle $\theta_{VB}$ with $|\sin \theta_{VB}| =0$. Therefore, outside this vicinity the cascade characteristic time scale $\tau_{cas} \geq \tau_{\perp}$. Application of the above-described stability principle (in respect to the Landau instability) results in $\tau_{cas} = \tau_{\perp}$. At the $\lim |\sin\theta_{VB}| \to 0$: $\tau_{\perp} \to 0$, i.e., the effective nonlinear interaction time becomes extremely fast. A cascade step cannot be taken before the information about the completeness of the sufficient non-linear interaction is spread over the entire wave packet. However, there is a causal restriction on the speed of the information propagation. This speed cannot be larger than the Alfv\'enic velocity, i.e., this limit is determined by $\tau_A$, i.e., $\tau_{cas} \geq \tau_A$. Therefore in this narrow vicinity of the angle $\theta_{VB}$: $\tau_{cas} \approx \tau_A$.\\
     
       While outside this angle vicinity, we can readily obtain the magnetic energy spectrum using the minimalistic model of the total energy $\mathcal{E}_{\ell} \propto v^2 \propto b^2$ cascade
\begin{equation} 
 \varepsilon \sim  \frac{\mathcal{E}_{\ell}}{\tau_{cas}} \sim  \frac{\mathcal{E}_{\ell}^{3/2}}{\ell |\sin\theta_{VB}|} \implies \mathcal{E}_{\ell} \sim (\varepsilon |\sin\theta_{VB}|)^{2/3} \ell^{2/3} \implies E(k) \sim  (\varepsilon |\sin\theta_{VB}|)^{2/3} k^{-5/3},
\end{equation}
inside the vicinity, the total energy flux can be estimated as follows
\begin{equation} \varepsilon \sim \frac{\mathcal{E}_{\ell}}{\tau_A} \sim \frac{\mathcal{E}_{\ell}V_A}{ \ell } \implies \mathcal{E}_{\ell} \sim  \frac{\varepsilon}{V_A}  \ell  \implies E(k)k \sim   \frac{\varepsilon}{V_A} k^{-1} \implies E(k) \sim \frac{\varepsilon}{V_A} k^{-2}. 
\end{equation}

  Let us now estimate the width of the tiny vicinity where the spectrum $E(k) \propto (\varepsilon |\sin\theta_{VB}|)^{2/3} k^{-5/3}$ is replaced by the spectrum $E(k) \sim \frac{\varepsilon}{V_A} k^{-2}$. The small critical angle $\theta_{VB}$ can be estimate from the equation $\tau_{\perp}=\tau_{\parallel}$:
 \begin{equation}
 \frac{\ell |\sin \theta_{VB}|}{v_{\ell}} \approx \frac{\ell |\cos\theta_{VB}|}{V_A} \implies \theta_{VB} \approx \frac{(\varepsilon \ell)^{1/2}}{V_A^{3/2}}
 \end{equation}
   
  The smallness of the causal angle is determined by the relative smallness of $\varepsilon \ell$. The smallness of this parameter is determined by three factors: 1) the smallness of the energy fluctuations in the magneto-inertial range of scales with respect to the large $V_A$ (which, in turn, is determined by the strong mean magnetic field), 2) the smallness of $\ell$ in this range. \\
  
  While the effective $\varepsilon$ is independent on $\ell$, it can depend on $\theta_{VB}$: in the imbalanced Alfv\'enic solar wind (with the different effective number of the outward and inward AWPs) $\varepsilon$ is decreased with $|\sin \theta_{VB}|$, that strengthens the geometrical trend which we have already partially taken into account. This results in an interesting phenomenon. Although in the imbalanced Alfv\'enic solar wind the number of the spatial domains with small $|\sin \theta_{VB}|$ should dominate, the individual energy of these domains is considerably decreased with $|\sin \theta_{VB}|$, and this decrease is mainly due to the same reason which leads to their dominance -- the anisotropic imbalance. Therefore, the global average (over all spatial domains) of the energy fluctuations can result in the spectrum $E(k) \sim  (\varepsilon |\sin\theta_{VB}|)^{2/3} k^{-5/3}$ and not in the spectrum $E(k) \sim \frac{\varepsilon}{V_A} k^{-2}$. The standard Taylor's hypothesis produces this average, and to observe the latter spectrum we should make a specific sampling of the observed time series to separate the intervals with small $|\sin \theta_{VB}|$. \\
   
   Both these spectra were observed in the solar wind \cite{hor},\cite{wick1}. In the paper \cite{hor}, the authors show that for $\theta_{VB}$ close to $0$ the observed in the magneto inertial range magnetic spectrum $E(k) \propto k^{-2}$, whereas for large $\theta_{VB} < \pi/2$ (where most of the magnetic power is contained)  the observed magnetic spectrum $E(k) \propto k^{-5/3}$. During the time of the measurements, the Ulysses spacecraft was within the fast solar wind at $R \approx 1.4$ AU. The data were processed using a wavelet method sensitive to the changing local mean magnetic field direction. The magnetic field fluctuations in the magneto-inertial range were relatively large; therefore, the causality cone was not so narrow as for the really weak turbulence.\\
  
  In paper \cite{wick1}, the authors select a fixed scale (or frequency $f$) well within the magneto-inertial range. Then, they extract the power $P(f, \theta_{VB})$ across all sampled angles from $0$ to $\pi$. Finally, they plot the normalized power $P(\theta_{VB}) / P(\pi/2)$against $\theta _{VB}$.  Spacecraft observations (from Ulysses and Wind missions) show that the power behaves like a smooth, sinusoidal-like distribution that matches the geometric reduction of the active cascade volume. 
  
\begin{figure} \vspace{-1cm}\centering \hspace{-0.7cm}
\epsfig{width=.6\textwidth,file=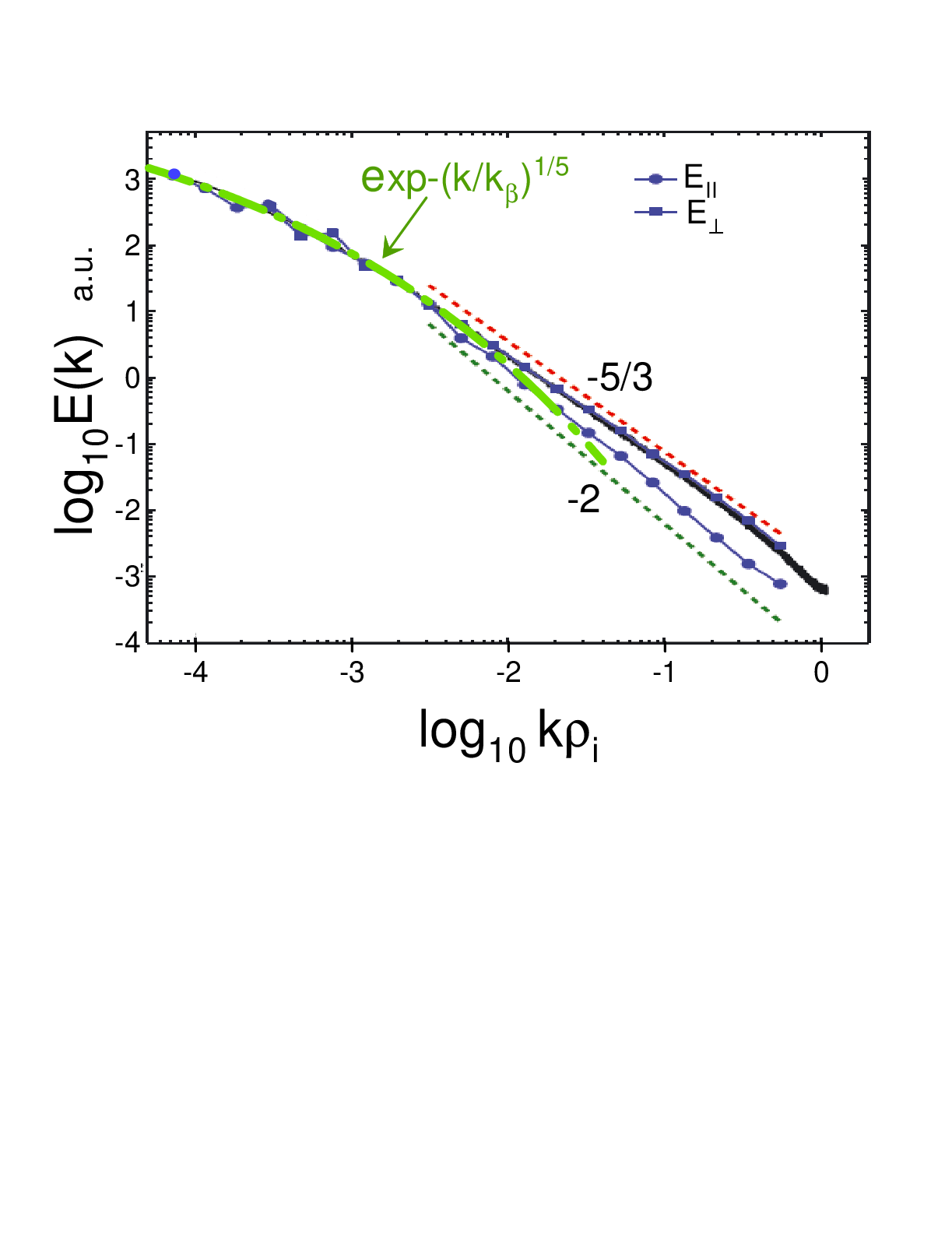} \vspace{-5.5cm}
\caption{Trace of the magnetic field power spectrum at $1. 48 < R < 2.11$ AU (Ulysses observations), $\rho_i$ is the ion gyroradius.} 
\end{figure}

  A clear separation of the two spectra $E(k) \sim k^{-5/3}$ and $E(k) \sim k^{-2}$ can be seen in Fig. 1, where the observed trace magnetic spectra obtained at $\theta_{VB}$ close to $\pi/2$ (top) and at $\theta_{VB}$ close to $0$ (bottom) are shown. The spectral data were taken from Fig. 1 of the paper \cite{wick2}. The Ulysses mission's raw data were obtained in the fast solar wind at $1.48 < R < 2.11$ AU and conditionally sampled to obtain the angularly determined spectra. The dashed curve indicates the stretched exponential large-scale spectrum $E(k) \propto \exp-(k/k_{\beta})^{1/5}$ corresponding to distributed chaos dominated by magnetic helicity \cite{b1} (its relation to the so-called reservoir of the Alfv\'enic wind will be discussed below in more detail).

\section{Renormalization in the magneto-inertial range: stochastic causality}

  Now let us consider the situation outside the tiny vicinity of the angle $\theta_{VB}$ with $|\sin\theta_{VB}| = 0$ within the magneto-inertial range in more detail, accounting for the stochastic nature of the process. In this case, the weak interactions of the contrapropagating Alfv\'enic wave packets in the magneto-inertial range can considerably slow down the total energy cascade. The slowdown is caused by two different phenomena: the short temporal window of the interactions (as in the classic IK approach) and the strong asymmetry between the contrapropagating AWP (imbalance). The physical workload required to complete a cascade step requires cross-field fluid shearing over the distance $\ell _{\perp }$. This shearing cannot happen arbitrarily; it is causally regulated by the parallel wave-packet crossings $\tau _{\parallel }$. Because each crossing only provides a weak, fractional update, causality dictates that plasma cannot transfer the energy out of the scale without executing at least enough parallel crossings to match the perpendicular fluid clock: 
$N_{cas}\ge \tau _{\perp }/\tau _{\parallel }$. 

   Just as in the hydrodynamic case, the system is subjected to severe external flux perturbations (e.g., from the expanding solar wind or macroscale shear streams). These perturbations drive localized accumulations of energy. If the system tries to cascade using an inefficiently high number of steps $N_{cas} > \tau_{\perp} / \tau_{\parallel}$, the total renormalized cascade time stretches unnecessarily long. The slow cascade fails to "drain" the localized energy flux accumulations fast enough. The localized spatial gradients --governed by the nonlinear $ (\nabla \times {\bf u} \times {\bf b})$  term -- will spike, rendering that state unstable to energy flux perturbations. To survive these perturbations and achieve structural stability, the system must choose the configuration that handles the flux with the highest possible efficiency. It drives the interaction count down to its minimum allowed causal threshold: 
\begin{equation}
 N_{cas}=\frac{\tau _{\perp }}{\tau _{\parallel }}.
\end{equation}  
 This choice is the exact magnetohydrodynamic analogue of the above-suggested hydrodynamic causality-stability criterion, enforced by the exact same physical requirement: the necessity to clear localized energy flux accumulations fast enough to maintain cascade stability.\\

  Let us consider the slowdown of the total energy cascade by the weak Alfv\'enic wave-packet interactions as a renormalization of the $\tau_{cas}$ by a stochastic renewal process with the binary distribution of the inter-renormalization waiting time $\tau$: 
 \begin{equation} 
 \psi(\tau) = p  \delta(\tau - \tau_{\parallel}) + (1-p) \delta (\tau - \tau_{\perp})
\end{equation} 
where $ \delta(x)$ is the delta function of $x$, $p$ and $(1-p)$ are the probabilities, $ 0 < p < 1$. With such a distribution, we firmly stand on the minimalistic floor. The mean waiting time 
 \begin{equation} 
  \langle \tau \rangle = \int \tau \psi(\tau) d\tau = p \tau_{\parallel} + (1-p) \tau_{\perp}.
\end{equation}

For $t \gg \langle \tau \rangle$, the expected number of renewals for the time $t$ is $\langle N \rangle \propto t/\langle \tau \rangle$ (the Elementary Renewal Theorem). 
Therefore, the mean number of the renewal steps per the renormalized cascade’s waiting time ${\tilde\tau}_{cas}$ is 
\begin{equation}
\langle N \rangle_{cas} \sim {\tilde\tau}_{cas}/ \langle \tau \rangle.
\end{equation}
  
    The number of events (wave packet crossings) necessary for one cascade step $N_{cas}$ cannot be changed by time rescaling. Therefore $\langle N \rangle_{cas} = N_{cas}$ and, consequently:
 \begin{equation}
\frac{\tau _{\perp }}{\tau _{\parallel}} = \frac{\tilde{\tau}_{cas}}{\langle \tau {\rangle}}
 \end{equation} 
 This relationship is aligned with the causality-stability principle. The left-hand side of this equation represents microscopic structural capacity, whereas the right-hand side represents the coarse-grained footprint. Substituting $\langle \tau \rangle$ in this equation we obtain

\begin{equation} 
 \tilde{\tau}_{cas} = \tau_{\perp}\{p +(1-p) \frac{\tau_{\perp}}{\tau_{\parallel}}\}
 \end{equation}

  Since outside the tiny vicinity of the angle $\theta_{VB} =0$ (see the estimate of this vicinity below): $\tau_{\perp} \gg \tau_{\parallel}$, for most of values of $p$ we can write 
 \begin{equation} 
 \tilde{\tau}_{cas} = (1-p) \frac{\tau_{\perp}^2}{\tau_{\parallel}}
 \end{equation} 
  
  excluding only the tiny vicinity  $\Delta p$ of $p=1$: 
\begin{equation}
 \Delta p \sim \frac{v}{V_A} \frac{|\cos \theta_{VB}|}{|\sin \theta_{VB}|}.
 \end{equation}
  Therefore, for the statistically robust solution ${\tilde \tau}_{cas} \sim \tau_{\perp} (\tau_{\perp}/\tau_{\parallel}) = \tau_{\perp}^2 / \tau_{\parallel} \gg \tau_{\perp}$ Analogously, for the limit $p \to 1$ (statistically marginal solution) we obtain $\tilde{\tau}_{cas} \sim \tau_{\perp}$.\\ 
   
  A).  For the statistically robust solution:
   
\begin{equation}
\varepsilon \sim \frac{  v^2}{\tilde{\tau}_{cas}}\sim \frac{  v^2}{(1-p) \tau_{\perp}^2/\tau_{\parallel}} 
\end{equation}
 Hence,
\begin{equation}
(1-p) \varepsilon \sim \frac{  v^2}{(\ell_{\perp}^2/  v^2)(\ell_{\parallel}/V_A)^{-1}} \sim \frac{\ell_{\parallel}  v^4}{\ell_{\perp}^2 V_A} \sim \frac{|\cos \theta_{VB}|  v^4}{\ell|\sin \theta_{VB}|^2 V_A}
\end{equation}

Then (taking into account $k \propto 1/ \ell$), 
\begin{equation}
  v^2 \sim  \frac{|\sin \theta_{VB}|}{\sqrt{|\cos \theta_{VB}|}}   ((1-p)\varepsilon V_A)^{1/2} \ell^{1/2} \implies E_v(k) \propto  \frac{|\sin \theta_{VB}|}{\sqrt{|\cos \theta_{VB}|}} ((1-p)\varepsilon V_A)^{1/2}k^{-3/2}.
\end{equation}

B). For the statistically marginal solution ($p \to 1$): the long pauses disappear, the bimodal clock becomes asymptotically deterministic, and ${\tilde \tau}_{cas} \to \tau_{\perp}= \ell_{\perp}/   v$.
This immediately yields:
\begin{equation}
\varepsilon \sim \frac{  v^2}{{\tilde \tau}_{cas}} \sim \frac{v^3}{\ell_{\perp}} \implies   v^2 \sim (\varepsilon |\sin \theta_{VB}|)^{2/3} \ell^{2/3} \implies E_v(k) \propto  (\varepsilon |\sin \theta_{VB}|)^{2/3}k^{-5/3}.
\end{equation}

  Using the relationship $v_{\ell}= C b_{\ell}$ we obtain analogous results for magnetic energy spectrum $E_b (k)$ (maybe with an intermittency correction which is beyond our present consideration, see above).\\

 \section{Boundaries of applicability and limit cases}
 
 Now let us check whether the conditions which were assumed for applicability of the stochastic renewal process can be satisfied by the obtained solutions. 
 
 For statistically robust solution: 
\begin{equation}
 \frac{\tilde{\tau}_{cas}}{\tau_{\perp}} \sim \frac{\tau_{\perp}}{\tau_{\parallel}} \sim \frac{|\sin \theta_{VB}|^{1/2}}{|\cos \theta_{VB}|^{3/4}}\frac{V_A^{3/4}}{(1-p)^{1/4} (\varepsilon \ell)^{1/4}} \gg 1.
\end{equation}
Outside a tiny vicinity of  $\theta_{VB} = 0$ this condition is satisfied due to weak turbulence condition 
\begin{equation}
\frac{ (\varepsilon \ell)^{1/4}}{ V_A^{3/4}} \ll 1.
\end{equation}

Let us estimate the order of the vicinity from 
\begin{equation} \tau_{\perp} \sim \tau_A \implies \theta_{VB}  \sim (1-p)^{1/2} \frac{(\varepsilon \ell)^{1/2}}{V_A^{3/2}} \ll 1.
\end{equation}
When $\tau_{\parallel} \geq \tau_{\perp}$ the causality condition $\tau_{cas} = \max \{\tau_{\perp},\tau_{\parallel}\}$ gives 
\begin{equation}
\tau_{cas}= \tau_{\parallel} \implies \varepsilon \sim \frac{v^2}{\ell_{\parallel}/V_A} \sim \frac{\varepsilon}{V_A} k^{-2}.
\end{equation}

 Now let us consider another limit $|\cos \theta_{VB}| \to 0$. At this limit the condition 
\begin{equation}
 \frac{\tilde{\tau}_{cas}}{\tau_{\perp}} \sim \frac{\tau_{\perp}}{\tau_{\parallel}} \sim \frac{|\sin \theta_{VB}|^{1/2}}{|\cos \theta_{VB}|^{3/4}}\frac{V_A^{3/4}}{(1-p)^{1/4} (\varepsilon \ell)^{1/4}} \gg 1.
\end{equation}
is perfectly satisfied, but $E_{v,b}(k) \to \infty$. That means an instability of the statistically robust solution A, when  $|\cos \theta_{VB}|$ approaches zero, and bifurcation into stable marginal solution B.\\
 
   This bifurcation is consistent with the large-amplitude, spherically-polarized, discontinuity-bearing character switchbacks are observed to have. Switchbacks can be a cause of the bifurcation, and the bifurcation can be a cause in situ for switchbacks. All these situations can have different properties. 
   
   Switchback $\rightarrow$ bifurcation: a large deflection is seeded elsewhere (coronal reconnection launches an already-folded structure), it propagates outward largely intact, and wherever it sits, the local turbulence it is embedded in is forced onto the marginal branch as a passive consequence -- the bifurcation describes the structure's local dressing, not its formation.
   
      Bifurcation $\rightarrow$ switchback: no external seed is needed at all; ambient turbulence alone, given enough amplitude and the right local field geometry, drives $\theta_{VB}$ past its own threshold spontaneously, and the marginal-branch state that appears is the switchback, generated in place.
 
   Both give the same local signature once formed. This is actually a natural explanation for why the existing literature looks so tangled (some studies find a clean SB/NSB spectral indixes for the switchbacks (-5/3) and ambient plasma (-3/2), others find none) \cite{bou,nay,mar,tat}: they may be sampling different mixtures of these two populations, with one being a genuine local generation event and the other one being a passively advected structure whose interior properties reflect coronal conditions more than local turbulence conditions.
   
   J. E. Borovsky \cite{bor} suggests switchbacks do not evolve after 0.3 AU, as if they had stopped interacting with the surrounding turbulent wave field, while other work argues the opposite, that near-Sun switchbacks do actively interact with the ambient field -- i.e., the field is already divided on whether the population looks "frozen" or "still coupled to local turbulence" (see, for a recent comprehensive review Ref. \cite{wyp}).

 \section{Dynamical phase transition in the magneto-inertial range}
  
    The statistically robust solution of the renormalization is inherently slow, while the statistically marginal solution is much faster and more efficient. The energy flux fluctuations can lead to the accumulation of energy in certain wavenumbers, which affects cascade stability (the above-discussed Landau instability). If the level of these fluctuations is subcritical, the slow cascade can successfully smooth (or ``transfer downward'') the localized accumulations. But if the level becomes supercritical, the system must switch to the faster (and more efficient) cascade -- the statistically marginal solution -- to flush out the energy accumulations before they can cause an instability. When the cause of the subcritical fluctuations of the energy flux ceases to exist, the system may again relax to the statistically robust state. However, if the cause is persistent (as, for instance, in the expanding solar wind), the fast solution becomes a permanent necessity. In the Alfvénic solar wind, different localized spatial regions can have different levels of energy flux fluctuations. Some of them can become supercritical while others cannot. The spatial ``drops'' of the supercritical behavior can appear in the subcritical ``sea'' and grow due to the solar wind expansion. It looks like a first-order phase transition and is generally consistent with observations.\\ 
    
  Let us describe the physical mechanism of the dynamical phase transition. The events of nonlinear wave packet interactions occur when wave packets moving in opposite directions collide. Statistically, this colliding process is most intensive in so-called balanced regimes, when the number of wave packets moving in opposite directions is approximately equal as well as their energies (these are the regions with $|\cos \theta_{VB}| \to 0$). Because nonlinear interactions are maximized in such balanced regimes, these localized domains are most conducive to generating energy flux fluctuations under environmental influence. The supercritical domains are more energetic than the subcritical ones and have much larger spatial gradients, because the corresponding nonlinear term in the MHD equations has a form $( \nabla \times [{\bf u} \times {\bf b}])$. At the same time, in the balanced regimes, the mean waiting time between these interactions is minimized. In the bimodal case, this corresponds to the limit $p \to 1$, representing the statistically marginal solution with the $k^{-5/3}$ spectrum. Hence, the very regions which are most conducive to generating energy flux fluctuations under environmental influence have the most effective mechanism for stabilizing the supercritical cascade (this is the physics behind pair of limits $p \to 1$ and $|\cos \theta_{VB}| \to 0$). \\ 
    
 In terms of dynamical systems theory, the subcritical $k^{-3/2}$ cascade operates as a structurally stable attractor with a wide basin of attraction. However, supercritical fluctuations fundamentally restructure the phase space by disrupting and destroying this slow attractor. In doing so, they transform the previously narrow basin of attraction of the marginal $k^{-5/3}$ cascade into a wide dominant one. Consequently, system trajectories are globally captured by this fast attractor, forcing the system into the only available physical state capable of smoothing or dissipating the intense energy flux. Notably, this topological restructuring is irreversible; once captured by the wide $k^{-5/3}$ basin, the system remains locked in this fast regime even if the initiating fluctuations subside. Besides, the main cause of the supercritical fluctuations -- the wind expansion -- is persistent; the fast solution becomes a permanent necessity in an increasing number of spatial locations. Taking into account that the supercritical regions are generally much more energetic than the subcritical ones, their contribution to the average energy, measured using the standard Taylor hypothesis, becomes progressively large with the wind expansion (see Fig. 1).

\section{Heterogeneous nucleation}

At heterogeneous nucleation a pre-existing seed lowers the barrier for a phase transition that would otherwise require a rare, spontaneous fluctuation.
The above described  $\theta_{VB}$-fold ($\cos \theta_{VB} \to 0$) is the natural seed for the Landau transition.

The two mechanisms differ in exactly the way that matters for nucleation theory: the $\theta_{VB}$ bifurcation is a hard, deterministic fold -- once a patch crosses it, there is no probabilistic element, the marginal branch is simply forced. The Landau/flux-fluctuation transition is soft and statistical -- it requires an local excursion of $\varepsilon$ to exceed the slow branch's relaxation capacity, which is exactly the situation classical nucleation theory describes as homogeneous nucleation: rare, because you need spontaneous fluctuations large enough to overcome an energetic barrier (here, the cost of maintaining a finite marginal-branch patch against relaxation pressure from the surrounding robust sea) with no help from a pre-existing interface. A switchback, regardless of what created it, is a location where that barrier has already been cleared deterministically, by geometry rather than by chance — it's a ready-made supercritical droplet, dropped into the sea rather than nucleated within it. That's precisely heterogeneous nucleation, and it's why heterogeneous nucleation is generically far faster and more common than homogeneous nucleation in real first-order transitions -- we don't need to wait for an unlikely fluctuation when a seed is already sitting there.\\

  The switchback boundary is a real, sharp current-sheet layer with enhanced local shear and wave activity. That boundary is exactly where we should expect elevated local $\varepsilon$ fluctuations -- the shear layer is a genuine local flux-intermittency source. So the growth mechanism is concrete: the seed's boundary locally elevates flux fluctuations in the immediately adjacent robust-branch material, pushing that material toward its own Landau threshold -- a real propagating front with the current sheet as the interface advancing into the subcritical sea.\\

  Therefore, switchbacks aren't caused by the bifurcation nor merely coincident with it -- they're the nucleation event that lets the broader statistical phase transition actually propagate through the plasma at a rate homogeneous nucleation alone couldn't achieve. The small $\theta_{VB}$ ($-5/3$) population becomes the signature of growth-front material -- robust-branch plasma freshly converted by proximity to a seed, not a spontaneous homogeneous-nucleation event.\\

\begin{figure} \vspace{-1cm}\centering \hspace{-0.7cm}
\epsfig{width=.56\textwidth,file=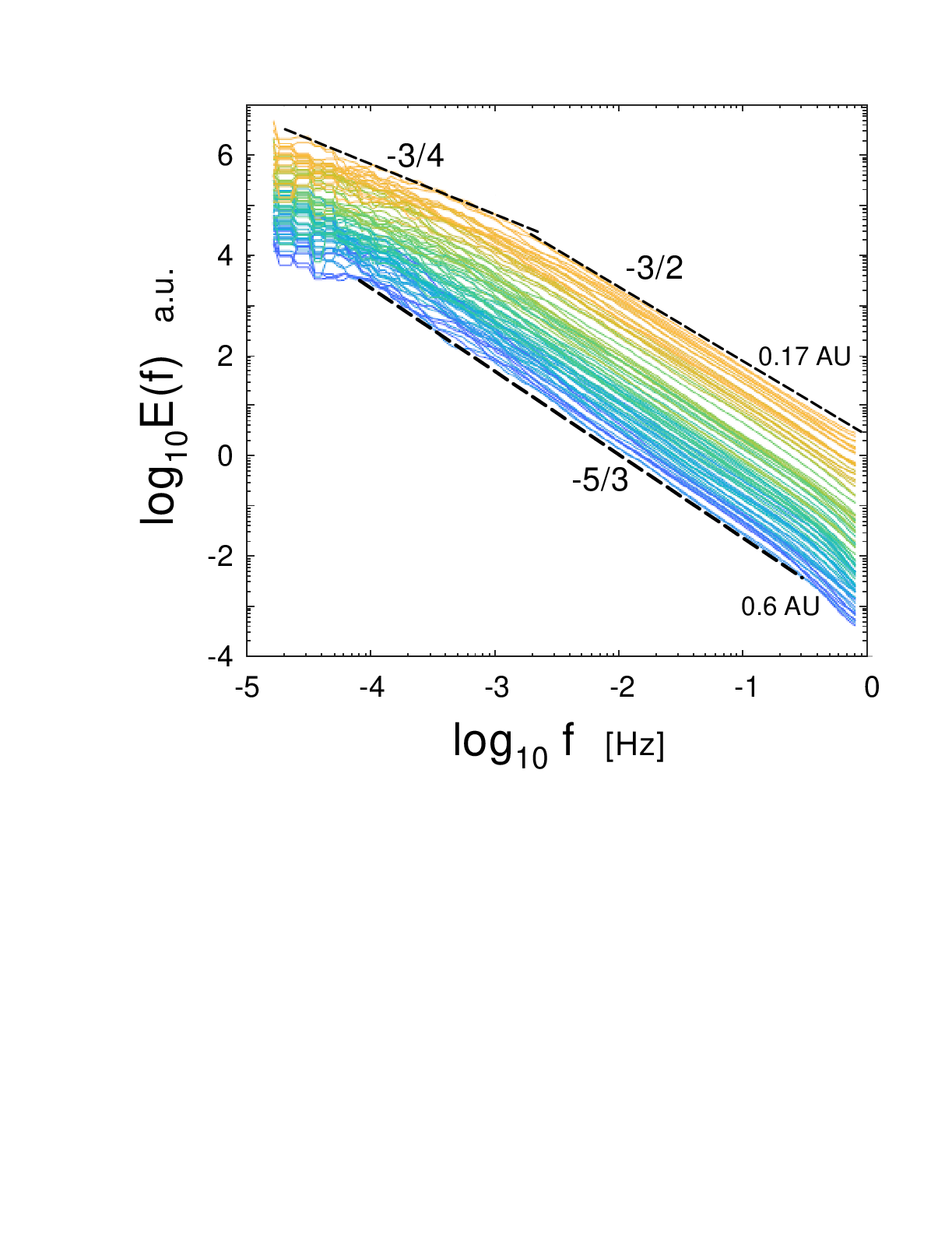} \vspace{-5cm}
\caption{Trace of the magnetic power spectrum from $R = 0.17$ AU to $R = 0.6$ AU (PSP first two orbits' observations).} 
\end{figure}

 Figure 2 shows the trace of the magnetic field power spectra from  $R = 0.17$ AU to $R = 0.6$ AU (Parker Solar Probe first two orbits' observations). The spectral data were taken from Fig. 1 of a paper \cite{chen}. In the inertial range of scales: for $R = 0. 17$ AU, the magnetic energy spectral index is $-3/2$, whereas for $R = 0.6$ AU, the magnetic spectral index is $-5/3$. The normalized cross helicity $\sigma_c = 2 \langle {\bf b} \cdot {\bf v} \rangle/ \langle {\bf b}^2 + {\bf v}^2 \rangle$ decays from $\sigma_c \approx 0.8$ to $\sigma_c \approx 0.3$ (see next Section).\\
  
Figure 3 shows the trace of the magnetic field power spectra for Alfvénic slow wind intervals at $R = 0.17$ AU (top, PSP observations), and at $R = 1$ AU (bottom, WIND observations). The spectral data were taken from Fig. 4 of a paper \cite{dpbv}. One can the the transition from the ``-3/2'' spectral index at $R = 0.17$ AU to ``-5/3'' spectral index at $R = 1$ AU in the inertial range of scales.
  
\begin{figure} \vspace{-0.7cm}\centering \hspace{-0.7cm}
\epsfig{width=.56\textwidth,file=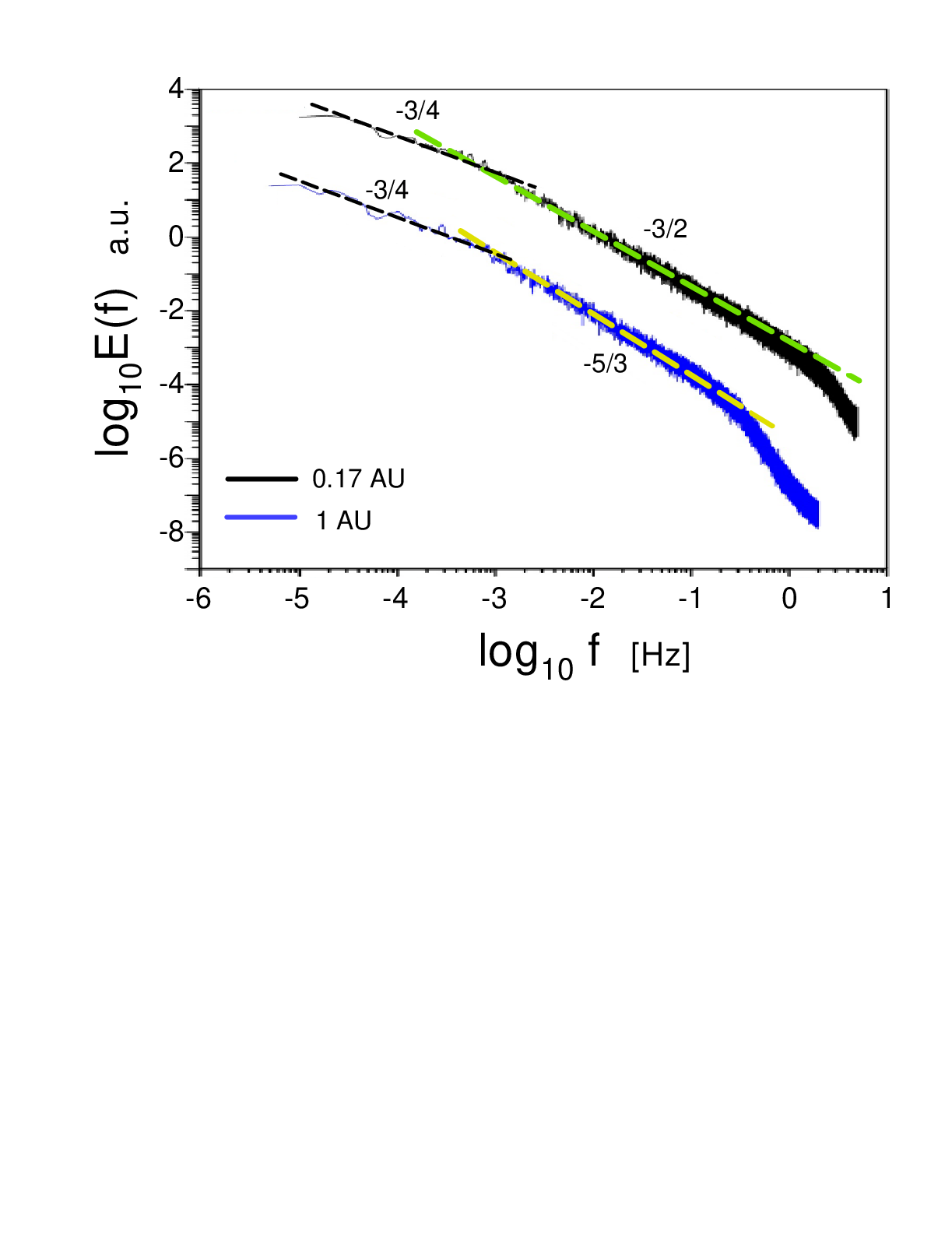} \vspace{-6cm}
\caption{Trace of the magnetic field power spectra for Alfvénic slow wind intervals at $R = 0.17$ AU (top, PSP observations), and at $R = 1$ AU (bottom, WIND observations).} 
\end{figure}
\section{Role of cross-helicity}
  
  In a genuinely Alfv\'enic solar wind, the limit $\sigma_c = 1$ can be approximately achieved when the local mean magnetic field is approximately parallel to the solar wind bulk velocity (radially oriented). In this case, the outward propagating wave packets have a solar origin, while the inward propagating wave packets are mostly in the minority. In the spatial domains with such orientation of the mean magnetic field, the statistically robust spectrum $E(k) \propto k^{-3/2}$ should be expected outside the tiny vicinity of $\theta_{VB} =0$. In the opposite case, when the local mean magnetic field is approximately perpendicular to the solar wind bulk velocity, the limit $ \sigma_c = 0$ is approximately achieved. The wave packets are moving along $\bf{B}$, and $\bf{B}$ is perpendicular to $\bf{V}$; then these wave packets are moving sideways (transversely) relative to the solar wind's expansion. They are not moving "inward" toward the Sun or "outward" away from it in a radial sense. Since radial solar injection typically cannot produce them, these balanced, transversely moving wave packets originate from particular local mechanisms. Let us consider one of the most prominent of them: the mechanism based on kinetic instabilities driven by cross-field currents. When $\bf{B} \perp \bf{V}$, the geometry naturally creates strong localized current sheets and magnetic gradients. These gradients trigger kinetic instabilities, such as drift or firehose/mirror instabilities, if the plasma pressure is anisotropic. These instabilities act as micro-explosions along the perpendicular field line, shedding fluctuations symmetrically to the left and to the right along $\mathbf{B}$, resulting in a perfect balance of wave energy. There are also many other local mechanisms. All this makes these domains very non-linear active, i.e., most susceptible to supercritical behavior and simultaneously the most favorable for the statistically marginal solution with the $k^{-5/3}$ spectrum (see above).  Notably, simultaneously with the above-described transition from the $k^{-3/2}$ spectrum to $k^{-5/3}$ spectrum, radial decay of the global normalized cross helicity $|\sigma_c|$ is observed as the solar wind expands. Such a correlation of the spectral index and $\sigma_c$ was previously mentioned for numerous observations (see, for instance, Ref. \cite{bb} and references therein). This global decay is exactly what occurs when the volumetric ratio of the highly energetic, balanced ($\sigma_c \approx 0$) drops expands, and increasingly populates the subcritical sea.

\section{Artificial decoupling of the spectral indices of velocity and magnetic field}

 Nature seems to be "economically" efficient at solving the cascade's stability problem. Why go to the total extreme reconstruction when the partial (preferably magnetic) reconstruction is sufficient to solve the emerging stability problem? Because the fluctuations are predominantly magnetic, the localized accumulations of supercritical energy are overwhelmingly concentrated in the magnetic field component. In the context of the model, the intense gradients triggering the transition to $p \to 1$ belong mainly to the magnetic field. Restructuring the velocity field requires moving bulk ion mass, which possesses significant fluid inertia. In contrast, the magnetic field can reconfigure its topology and sharpen its gradients far more rapidly through electron and ion kinetic micro-instabilities without needing to accelerate the entire bulk fluid to a new turbulent state. Both magnetic and velocity power spectra in the supercritical domains are transformed from the $k^{-3/2}$ into the $k^{-5/3}$ regime, but magnetic energy in the supercritical domains is considerably larger than the kinetic energy. Therefore, the contribution of the supercritical domains to the global magnetic energy is considerably larger than the corresponding contribution to the global kinetic energy. \\
 
  And indeed,  numerous observations in the inner Heliosphere show that the global (averaged over the sub- and super-critical domains) velocity spectrum, obtained using the standard non-sampling analysis of the measured time series, mixing the sub- and super-critical intervals) stays closer to $k^{-3/2}$ while the global magnetic spectrum transitions to $k^{-5/3}$ (see, for instance, \cite{bow}, and references therein). It is also found in Ref. \cite{bow}, using observations from the Wind spacecraft, that this persistent phenomenon is associated with magnetically dominated intervals in the measured time series (strongly magnetically dominated super-critical domains). A statistical analysis of the normalized cross helicity in the same observations shows that the existence of this phenomenon is strongly correlated with low cross helicity values (the balanced regime).  The authors of the paper also concluded that the steepening of the magnetic field to $-5/3$ is driven specifically by the formation of localized, highly energetic magnetic structures, which leave the velocity field completely unaffected. These observations are in good agreement with the phase transition approach.
  
  As observed by the Voyager probes, once the solar wind travels far enough from the Sun, the relative number of $-5/3$ patches approaches $100\%$ (first reported in a Ref. \cite{rob}). At this point, the $-3/2$ background patches are practically gone. The normalised cross-helicity $\sigma_c$ becomes low (cf above).  When the patchwork becomes uniform, the globally averaged velocity spectrum finally steepens to match the magnetic spectrum at a universal $-5/3$ slope.

\section{Magnetic helicity in solar wind}

\subsection{Magnetic helicty presence}

   Magnetic helicity is abundantly injected into the solar wind by the solar atmosphere. What people actually use in solar wind turbulence is the helicity of the fluctuating field $\mathbf{b} $, computed relative to the mean field. This quantity is well-defined and behaves as an ideal invariant of the fluctuation dynamics — it's one of the three quadratic invariants of incompressible MHD (total energy, cross helicity, magnetic helicity \cite{mt}) that survive in reduced/strong-guide-field MHD, where the dominant nonlinear couplings are quasi-2D in the plane transverse to the local mean magnetic $\mathbf{B}$. So even though the global helicity is ill-posed, the turbulence-relevant helicity with density  $h =\langle {\bf a} \cdot {\bf b} \rangle$  (where ${\bf b} = \nabla \times {\bf a}$) is perfectly sensible and still constrains cascade dynamic (helicity cascades/decays more slowly than energy), which shapes the spectral slopes and intermittency properties actually observed.

  It tags coherent structures and traces solar origin. Discrete structures embedded in the wind -- flux ropes, magnetic clouds within ICMEs, and (in some analyses) switchbacks -- carry a local twist/chirality whose sign is a helicity signature. Because coronal magnetic helicity injection follows statistical hemispheric preferences (the "hemispheric helicity rule"), the chirality observed in interplanetary flux ropes can be linked back to their solar source region, making helicity a diagnostic tool for solar–heliospheric connectivity and for testing flux-rope/CME models.

   Magnetic helicity can play a crucial role in the magneto-inertial range of some Alfv\'enic  solar winds (Bershadskii 2024) and (as will be shown below) at the largest (reservoir's) scales of Alfv\'enic  winds.

 Old standard 3D MHD phenomenology treats magnetic helicity as the more robust invariant -- lower-order in gradients than energy dissipation, which typically wants to pile up at large scales via an inverse cascade, not stream forward through a local, scale-by-scale flux the way energy does. Direct numerical simulations, however, show that if magnetic helicity is injected at a wavenumber $k_i$, then for $k< k_i$ there is an inverse cascade of magnetic helicity, whereas for $ k > k_i$ the magnetic helicity cascades downward.
 
   If magnetic helicity is injected at the largest scales (as it occurs in the reservoir of Alfv\'enic  winds), there is practically no room for the favored inverse cascade, and, therefore, magnetic helicity must perform an unfavored forward cascade. Since the topological ``charge'' related to magnetic helicity cannot be easily discharged, the system chooses the least energy-consuming way of mere stretching/advection (suggested by the induction equation) of the hard-to-destroy scalar with an independent of k magnetic helicity flux $\varepsilon_h$ mainly determined by the solar injection. It is much easier to adjust the magnetic field dynamics for this purpose inside the patches than to change its topology. Topological changes occur mainly via reconnections at the patch boundary, which consists of current sheets.\\

\subsection{A minimalistic model of dual cascade}

A statistically robust minimalistic model for such a cascade can be composed from a closed system of two algebraic equations for two variables $v_{\ell}$ and $b_{\ell}$: 
\begin{equation}
\varepsilon \sim \frac{(v_{\ell}^2 +b_{\ell}^2)}{\tau_{\perp}^2/\tau_{\parallel}} ~~ \text{and} ~~ \varepsilon_h  \sim \frac{h}{\tau_{\perp}} \sim \frac{\ell_{\perp} b_{\ell}^2}{\ell_{\perp}/v_{\ell}} 
\end{equation}
Hence
\begin{equation}
(\varepsilon V_A)\ell \sim v_{\ell}^4  + \varepsilon_h v_{\ell} ~~~\text{and} ~~~ b_{\ell}^2 \sim \varepsilon_h/v_{\ell}.
\end{equation}

The trigonometric prefactors are not used here for simplicity of the presentation. Notably, the cascade time used in the equation for the magnetic helicity cascade is taken $\tau_{\perp}$ and not $\tau_{\perp}^2/\tau_{\parallel}$ (as for the total energy cascade) because the system chooses the least energy-consuming way of mere stretching/advection (suggested by the induction equation). This algebraic system of equations has one real positive solution in standard radicals (rather cumbersome to be written here), with a scaling asymptotic 
\begin{equation}
v_{\ell}  \sim (\varepsilon V_A)^{1/4} \ell^{1/4}
\end{equation}
and 
\begin{equation}
b_{\ell} \sim \varepsilon_h^{1/2} (\varepsilon V_A)^{-1/8} \ell^{-1/8}
\end{equation}
at $\ell \gg \varepsilon_h^{4/3} (\varepsilon V_A)^{-1}$ (this asymptotic implies $v_{\ell}^2 \gg b_{\ell}^2$). Corresponding energy spectra are: 
\begin{equation}
E_v(k) \sim (\varepsilon V_A)^{1/2} k^{-3/2},
\end{equation}
and 
\begin{equation}
E_b(k) \sim \varepsilon_h (\varepsilon V_A)^{-1/4} k^{-3/4}.
\end{equation}

   Statistically marginal version of the model 
\begin{equation}
 \varepsilon \sim \frac{(v_{\ell}^2 +b_{\ell}^2)}{\tau_{\perp}} \implies (\varepsilon \ell) \sim v_{\ell}^3  + \varepsilon_h 
\end{equation}
 and 
\begin{equation}
\varepsilon_h  \sim \frac{h}{\tau_{\perp}} \sim \frac{\ell_{\perp} b_{\ell}^2}{\ell_{\perp}/v_{\ell}} \implies b^2 \sim \varepsilon_h/v_{\ell}.
\end{equation}
This system also has one real positive solution with a scaling asymptotic 
\begin{equation}
v_{\ell} \sim \varepsilon^{1/3} \ell^{1/3}
\end{equation} 
and 
\begin{equation}
b_{\ell} \sim \varepsilon_h^{1/2} \varepsilon^{-1/6} \ell^{-1/6}.
\end{equation}
   
   Corresponding energy spectra are: 
\begin{equation}
E_v(k) \sim (\varepsilon)^{2/3} k^{-5/3},
\end{equation}
and 
\begin{equation}
E_b(k) \sim \varepsilon_h (\varepsilon)^{-1/3} k^{-2/3}.
\end{equation}
\\
   
    Because the data in the reservoir large-scale range is messy and short, scientists from the very beginning used "hand-waving" explanations to simply round the observed spectral exponent to a neat, easily explainable whole number (-1). Since then, it has become clear that the reservoir spectrum is usually much flatter (see, for instance,  recent Refs. \cite{wu, sub} and references therein).  \\
    
    Figures 2 and 3, for instance, show a ``-3/4'' magnetic spectral index (cf Eq. 28).  Figure 4 shows an example of a trace magnetic spectrum measured by Helios 2 at $R =0.29$ AU in a fast solar wind observed in the ecliptic. The spectral data were taken from Fig 1 of Ref. \cite{tbt}.  In Ref. \cite{wu}, statistical analysis of the magnetic and velocity spectral exponents in the fast solar wind at $R = 1$ AU was performed, and ``$-0.78 \pm 0.22$'' exponent for magnetic energy and ``$-1.49 \pm 0.3''$ spectral exponent for corresponding velocity exponent were found in the reservoir range of scales (cf Eqs. (27-28)). A clear kinetic energy dominance was found deep in the reservoir range. The dominance decreases toward the breaking point.  Notably, the breaking point between the reservoir and magneto-inertial range is at $v_{\ell}^2 \sim b_{\ell}^2$.
 
\begin{figure} \vspace{-0.8cm}\centering \hspace{-0.7cm}
\epsfig{width=.48\textwidth,file=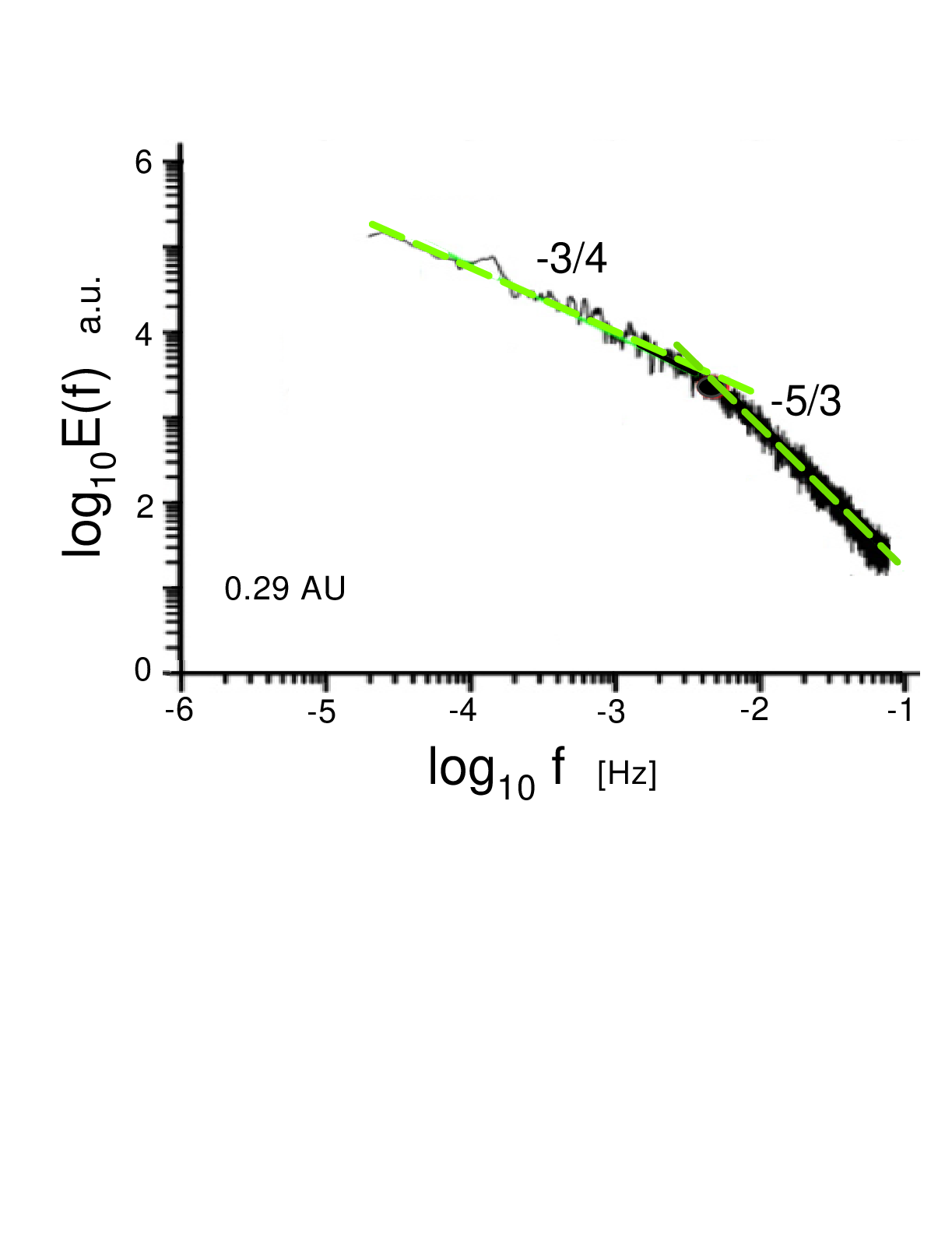} \vspace{-4.1cm}
\caption{Trace of the magnetic field power spectra in fast solar wind at $R = 0.29$ AU (Helios 2 observations in the ecliptic).} 
\end{figure}

\subsection{Distributed chaos dominated by magnetic helicity}
  
 With the replacement $\ell \sim 1/k$  a real-space amplitude scaling Eqs. (26) and (32) can be written as

\begin{equation}
b_{k} \sim \varepsilon_h^{1/2} (\varepsilon V_A)^{-1/8} k^{1/8}
\end{equation} 
\begin{equation}
b_k \sim \varepsilon_h^{1/2} \varepsilon^{-1/6} k^{1/6}.
\end{equation}    
seem very similar to the key Eqs. (17) and (16) of Ref. \cite{b1} devoted to distributed chaos in the magneto-inertial range of space plasma (not only the values of the exponent but also the dimentional prefactors are the same). It is not surprising because in Ref. \cite{b1} the idea of a passive-scalar-like behavior of magnetic helicity was also employed. The main difference occurs at the next step. Here we used an estimate $b_k^2 \sim E_b(k) k$ to obtain a scaling (power-law) spectrum, whereas in Ref. \cite{b1} a probabilistic approach was used to obtain the distributed chaos  (stretched exponential) spectrum. The two approaches therefore share the same underlying cascade-derived amplitude law and differ only in the final step by which that amplitude is converted into an observed spectral density: the standard scaling  substitution $b^2_k\sim E_b(k) k$ used above, or the probabilistic construction of Ref.~\cite{b1}, in which $b_k$ is instead treated as a stochastic variable and the distribution of $k$ is derived from the scaling relation
\begin{equation}
b_k \sim k^{\alpha} 
\end{equation}
  It is shown in Ref. \cite{b1} that if $b_k$ has a half-normal distribution with zero mean, the distributed chaos spectrum of the magnetic field has a stretch-exponential form
\begin{equation}
E_b(k) \sim \exp{-(k/k_{\beta})^{\beta}}.
\end{equation}
where the exponents $\alpha$ and $\beta$ are related 
\begin{equation}
\beta = \frac{2\alpha}{(1 + 2\alpha)}
\end{equation}
   
 Therefore, the spectral index ``-3/4'' for the scaling spectrum corresponds to the stretched exponential spectrum
\begin{equation}
E_b \sim \exp-(k/k_{\beta})^{1/5}
\end{equation}
for distributed chaos: $\alpha = 1/8 \implies \beta = 1/5$.  Analogously for the marginal case $\alpha = 1/6 \implies \beta = 1/4$.\\

  Notably, it is shown in the paper \cite{b1} that the statistically robust (in the present paper's terms) case $\beta =1/5$ is observed in the inner heliosphere, whereas the statistically marginal case $\beta =1/4$ is observed in the outer heliosphere. This finding aligns with the phase-transition concept presented in the present paper.\\

\begin{figure} \vspace{-1cm}\centering \hspace{-0.7cm}
\epsfig{width=.54\textwidth,file=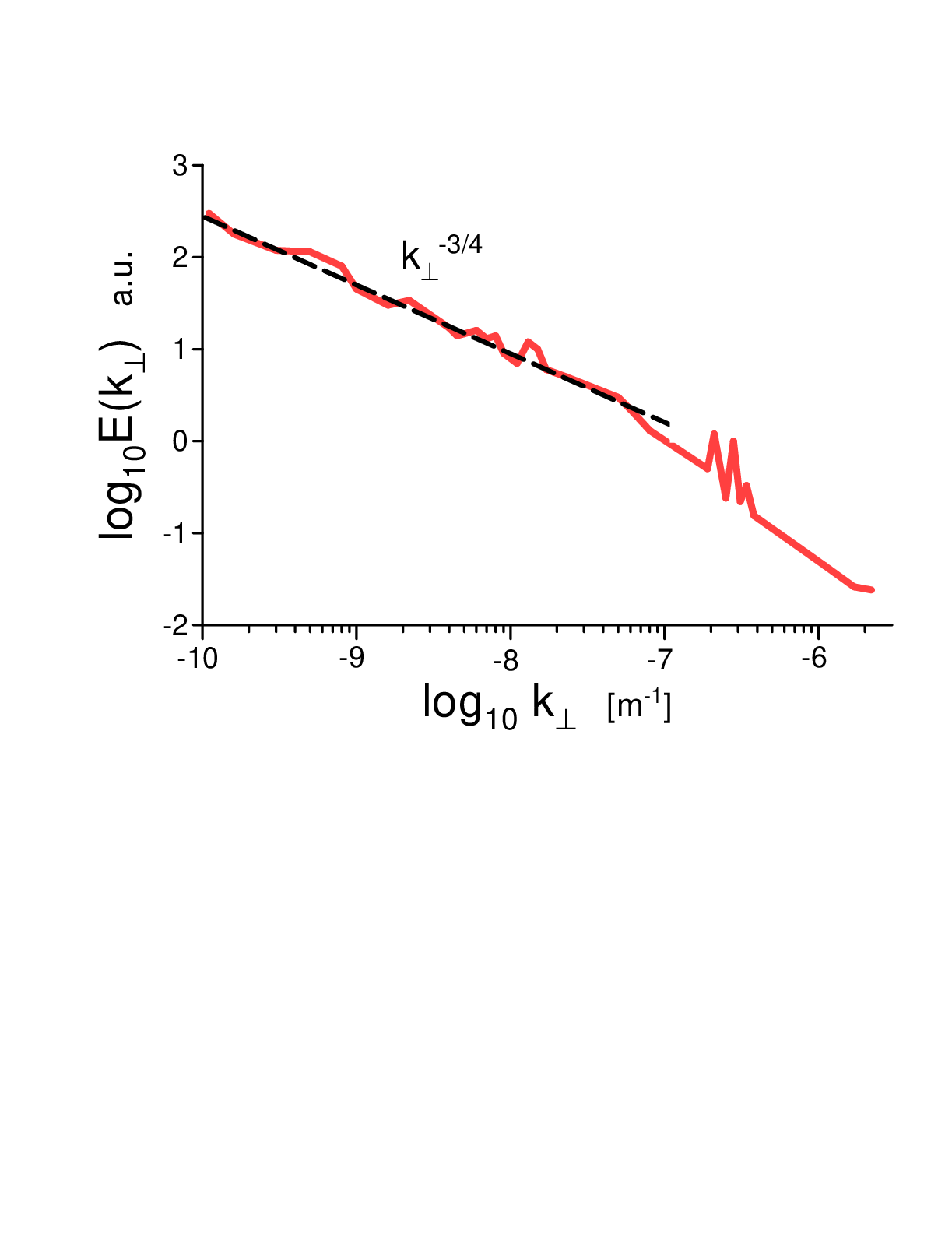} \vspace{-5.3cm}
\caption{Trace of the magnetic field power spectra in a pre-shock region of the outer heliosheath (Voyager 1 observations at $R = 140$ AU).} 
\end{figure}
\begin{figure} \vspace{-0.5cm}\centering \hspace{-0.7cm}
\epsfig{width=.56\textwidth,file=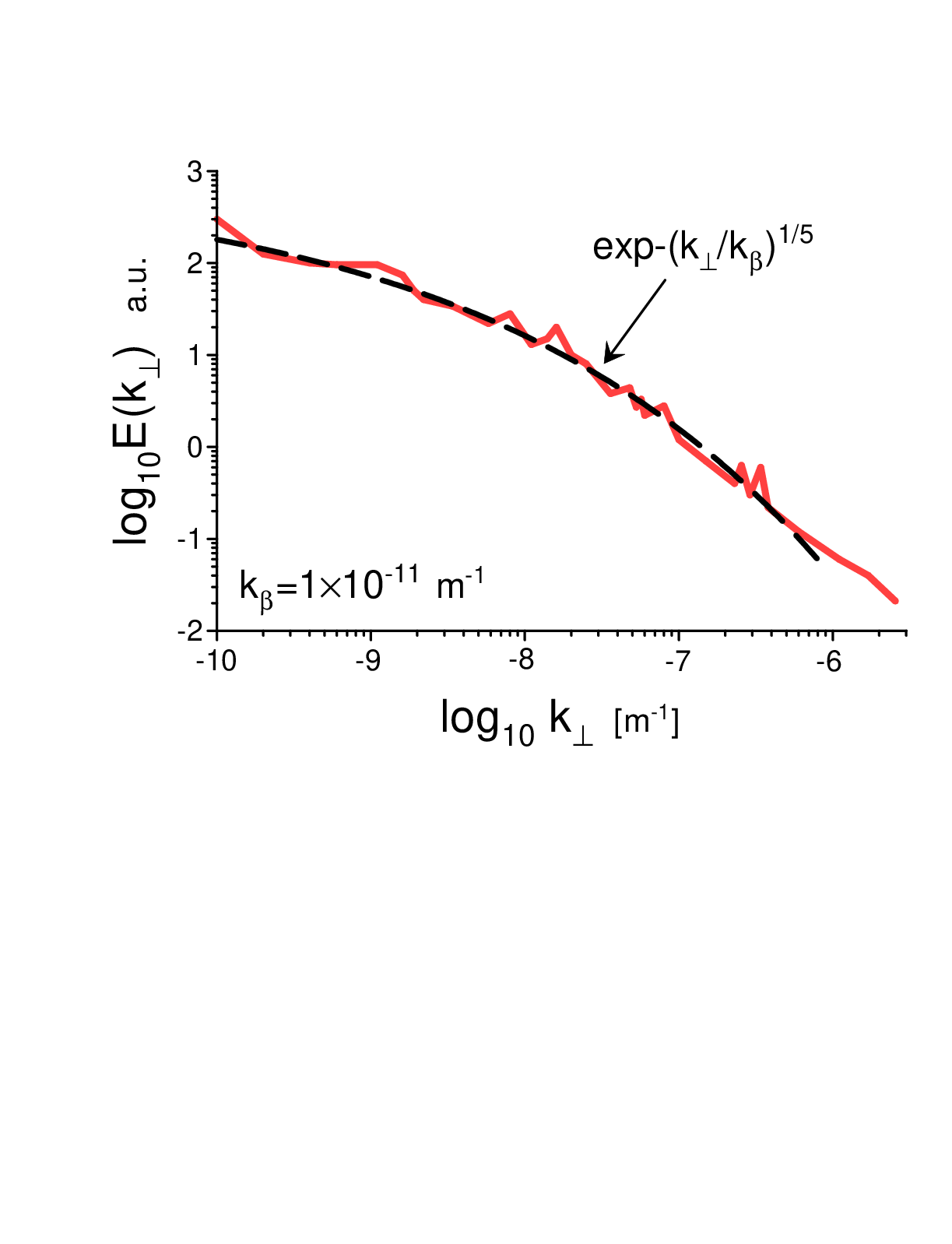} \vspace{-5.3cm}
\caption{Trace of the magnetic field power spectra in a post-shock region of the outer heliosheath (Voyager 1 observations at $R = 140$ AU.} 
\end{figure}

   The first relevant example of this phenomenon is given in Fig.1, where an extended reservoir of a fast wind is best fitted by Eq. (40) (the numerous observations of the distributed chaos spectra in the solar wind, magnetosphere of the Earth and planets, in the active solar regions, and in the interstellar plasma one can find in the Refs. \cite{b1},\cite{b3}). \\

  Another interesting example can be found in the outer heliosheath and is related to the propagation of a shock wave. Figures 5 and 6 show the traces of the magnetic field power spectra in the pre-shock (Fig. 5) and post-shock (Fig. 6) regions. The spectral data were taken from a paper \cite{fpb} and present Voyager 1 observations at $R = 140$ AU. Shock waves often cross the outer heliosheath and travel outward into the very local interstellar medium. The shock wave being examined propagates approximately radially. Rather than reflecting two distinct cascade physics, these two figures may reflect two different projections of a single, common cascade result onto the spectral domain: the spectral index -3/4 for the scaling spectrum (Fig. 5), and the stretched exponential spectrum with $\beta = 1/5$ for the distributed chaos (Fig. 6). The fundamental and tantalizing question of what physically determines which projection is realized in a given situation remains open for future investigations (see also Ref. \cite{mm} and references therein). For instance, the pre-shock spectrum (Fig. 5) is the scaling one and the post-shock spectrum (Fig. 6) is stretched-exponential. A reasonable reading is that the post-shock plasma has been compressed and processed into a smoother state (cf. Introduction section of Ref \cite{b1}).
  
\section{Switchbacks in the upper solar corona }

The Alfv\'en surface, where the wind speed equals the local Alfv\'en speed ($M_A =V_{sw}/V_A =1$), separates
the sub-Alfv\'enic corona from the super-Alfv\'enic wind. It is corrugated rather than
spherical, with a typical heliocentric distance of order $10$--$20$ $R_\odot$ that varies with solar
activity. PSP first crossed it in April 2021, and later perihelia show
repeated entries and exits, so the effective boundary is a zone a few to $\sim 10\, R_\odot$ thick
(order-of-magnitude estimates; measurements remain sparse).\\

  The switchbacks (which often contain most of the magnetic fluctuation energy) have been recently observed around the Alfv\'en surface \cite{huang1, huang2}. Since the switchbacks play a significant role in the solar wind proper (see above) it is interesting how they behave at the very beginning of the solar wind.
  It is shown above that the switchbacks are natural places for the marginal spectrum Eq. (17).
  
  The recent observational campaigns by the Parker Solar Probe (PSP) near the Alfv\'enic surface have revealed the existence of clean, extended frequency power-law magnetic spectra spanning up to two full decades with an apparent $f^{-5/3}$ spectral index in the magneto-inertial range and the large-scale scaling spectra shallower than $f^{-1}$ \cite{huang1,huang2}. It should be noted that not all spectra shown in the Refs. \cite{huang1,huang2} show such spectral behavior, which depends on the solar atmosphere dynamics.\\
  
  Inside the sub-Alfv\'enic region, the solar wind bulk velocity $\mathbf{V}_{sw}$, the intrinsic Alfv\'en wave propagation speed $\mathbf{V}_A$, and the spacecraft's orbital velocity $\mathbf{V}_{sc}$ are all of the same order of magnitude. This thaws the classical Taylor's frozen-in condition, complicating the space-to-time mapping. Inside the Alfv\'en surface, the spacecraft detectors do not sample the plasma along the radial direction, nor along the isolated vector $\mathbf{V}_{sc}$. Rather, they register fluctuations along the direction of the total relative advection velocity in the probe's frame of reference, defined by the effective velocity vector \cite{klein}:
\begin{equation}
\mathbf{V}_{eff} = \mathbf{V}_{sw} + \mathbf{V}_A - \mathbf{V}_{sc}
\label{eq:v_eff}
\end{equation}

  The measured wavenumber along the spacecraft's temporal timeline ($k_{obs}$) is therefore a geometric projection of the physical radial wavenumber ($k$) onto the direction of $\mathbf{V}_{eff}$:
\begin{equation}
k_{obs} = k \cos \phi_{eff}
\label{eq:k_projection}
\end{equation}
where $\phi_{eff}$ is the observational projection angle between the radial vector and $\mathbf{V}_{eff}$. 

Crucially, inside the Alfv\'en surface near perihelion, the radial solar wind speed is undergoing initial acceleration ($V_{sw,R} \sim 200$~km/s) and matches the radial Alfv\'en speed ($V_{A,R} \sim 200$~km/s), providing a total radial wave advection baseline of $\sim 400$~km/s. Concurrently, the spacecraft cuts sideways through the corona with a massive tangential speed ($V_{sc,T} \sim 200$~km/s). Because these bulk velocity components evolve smoothly over orbital timescales that are orders of magnitude longer than the rapid turbulent fluctuations, the ratio between the radial and tangential components of $\mathbf{V}_{eff}$ remains rigidly locked. Consequently, the projection angle $\phi_{eff}$ behaves as a \textit{quasi-stationary constant} ($\phi_{eff} \approx 26.5^\circ$) throughout the encounter window.  Notably, recent PSP observations generally validate the assumption made in the paper \cite{klein}.

Applying this linear coordinate transformation to the spectrum Eq. (17) yields:
\begin{equation}
E(k_{obs}) = E(k) \cdot \left| \frac{dk}{dk_{obs}} \right| \propto k_{obs}^{-5/3}
\label{eq:trans1}
\end{equation}

  This demonstrates that while the highly oblique projection onto $\mathbf{V}_{eff}$ introduces a stable geometric scaling factor to the absolute spectral density, it leaves the true $-5/3$ cascade exponent entirely untouched. Analogous consideration is valid also for the large-scale spectrum Eq. (34).\\
  
\begin{figure} \vspace{-0.5cm}\centering \hspace{-0.7cm}
\epsfig{width=.5\textwidth,file=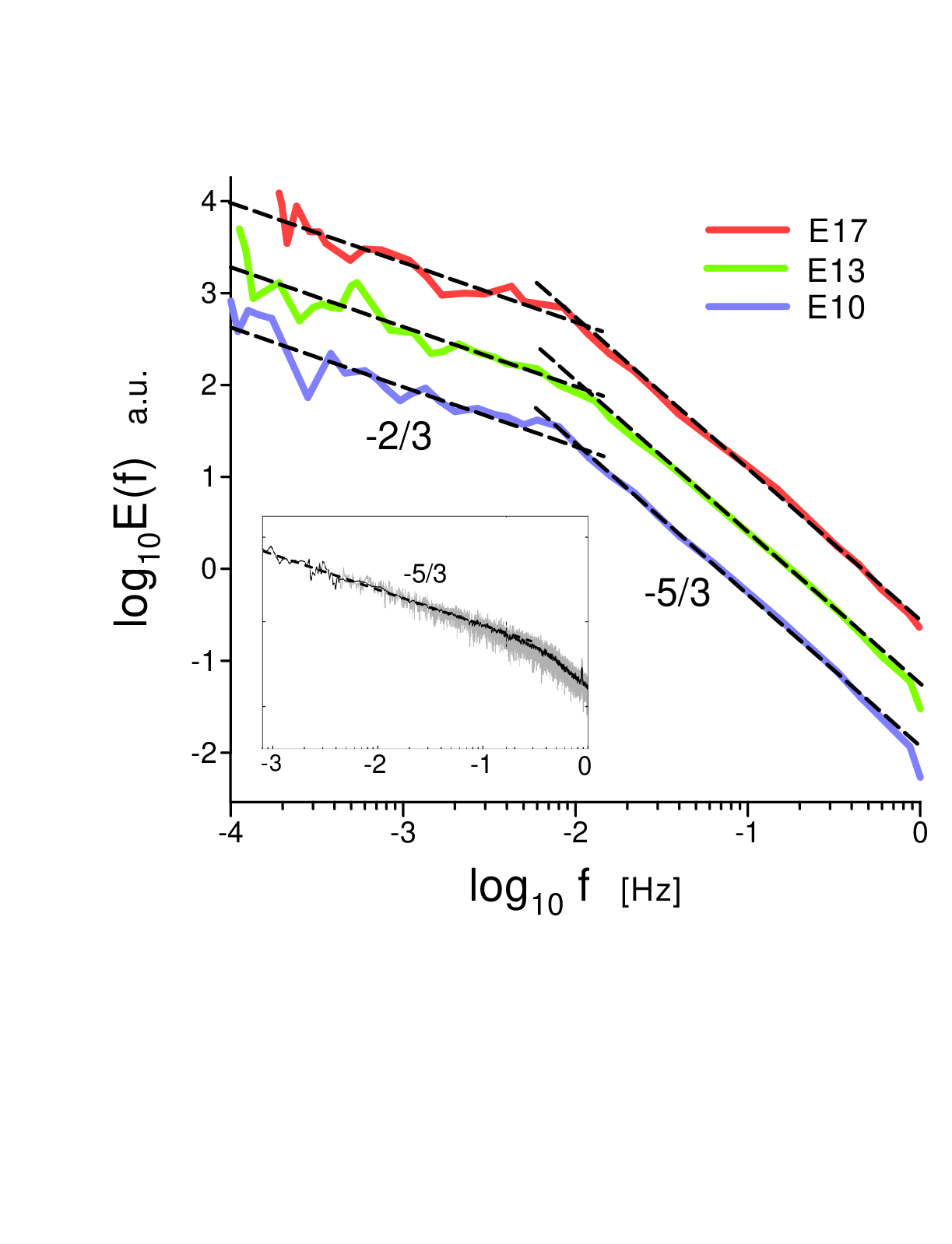} \vspace{-3.2cm}
\caption{Trace of the magnetic field power spectra for the switchback-dominated intervals in the upper solar corona (E17: at $R \approx 0.056$ AU). The insert shows the magnetic field power spectra for a typical switchback in the young solar wind at $R \approx 0.17$ AU. } 
\end{figure}

  Figure 7 shows the trace of the magnetic field power spectra for the (pre-) switchback-dominated intervals in the upper solar corona (for the E17 encounter $R \approx 0.056$ AU). The dashed straight lines indicate the power law Eq. (34) for large scales and Eq. (17) for the intermediate scales, respectively. The spectral data were taken from the Refs. \cite{huang1, huang2}. The insert shows the magnetic field power spectra for a typical switchback in the young solar wind at $R \approx 0.17$ AU for comparison. The spectral data for the insert were taken from Ref. \cite{sbb}.

\section{Astrophysical Plasmas}

\subsection{Milky Way}

 As shown in Fig. 5, the helical cascade in its scaling spectrum realization can become dominant in the outer heliosheath. 

   Supernova remnants and superbubbles (such as the Local Bubble boundary) in the local Milky Way interstellar mediumin inject kinetic and magnetic energy at outer scales of  $L_0 \sim 50\text{--}150\text{ pc} $. At this  $\sim 100\text{ pc} $ driving scale, Galactic differential rotation and vertical density stratification break mirror symmetry: as a superbubble expands across the disc's density gradient, the Coriolis force twists the rising plasma, injecting kinetic helicity that winds magnetic field lines into helical loops. Because total magnetic helicity is conserved, this process generates equal and opposite signs of magnetic helicity. Under large-scale  $\alpha\text{--}\Omega $ dynamo action, one sign undergoes an inverse cascade to sustain the coherent, kiloparsec-scale Galactic magnetic field, while the opposite sign remains at  $\sim 100\text{ pc} $ and cascades forward to smaller scales (or is vented into the halo via superbubble blowouts to avoid quenching the dynamo).
This two-regime framework accounts naturally for recent 3D Bayesian Faraday-rotation inversions of the local ISM. Using an observed full-sky Faraday depth map,
 \begin{equation}
 \phi =\frac{e^{3}}{2\pi m_{e}^{2}c^{4}}\int _{\text{LoS}}dl\,n_{e}B_{\parallel },
 \end{equation}
anchored to an independent 3D electron density ( $n_{e} $) distribution from dust-extinction-informed simulations (thereby decoupling magnetic fluctuations from thermal density structure), McCallum et al. [43] isolate the line-of-sight (LoS) magnetic field  $B_{\parallel } $ in 3D space rather than projecting it onto a 2D sky plane. Face-on reconstructions reveal coherent magnetic patches with sign reversals on  $\sim 100\text{ pc} $ scales and yield a 3D magnetic spectral index of  $n_{3\text{D}} = -2.73 \pm 0.19 \approx -11/4 $.\\

  Physically, these coherent  $\sim 100\text{ pc} $ structures act as the macroscopic guide-field domains ( $L \gg \ell $) of our framework. Within each patch, the local mean magnetic field regulates the Alfvénic timescale  $\tau _{A} $ and drives the direct helicity cascade in the outer-scale kinetic-energy-dominated regime ( $v_\ell^2 \gtrsim b_\ell^2 $, cf. Section IX.B). While each individual domain is locally anisotropic, averaging over an ensemble of randomly oriented patches across the inversion volume restores global statistical isotropy. Applying the standard isotropic-shell relation  $n_{3\text{D}} = n_{1\text{D}} - 2 $ maps the predicted 1D scaling  $n_{1\text{D}} = -3/4 $ (cf. Eq. (28)) directly onto  $n_{3\text{D}} = -11/4 = -2.75 $, matching the observed value within  $1\sigma $.\\
  
\begin{figure} \vspace{-1cm}\centering \hspace{-0.7cm}
\epsfig{width=.54\textwidth,file=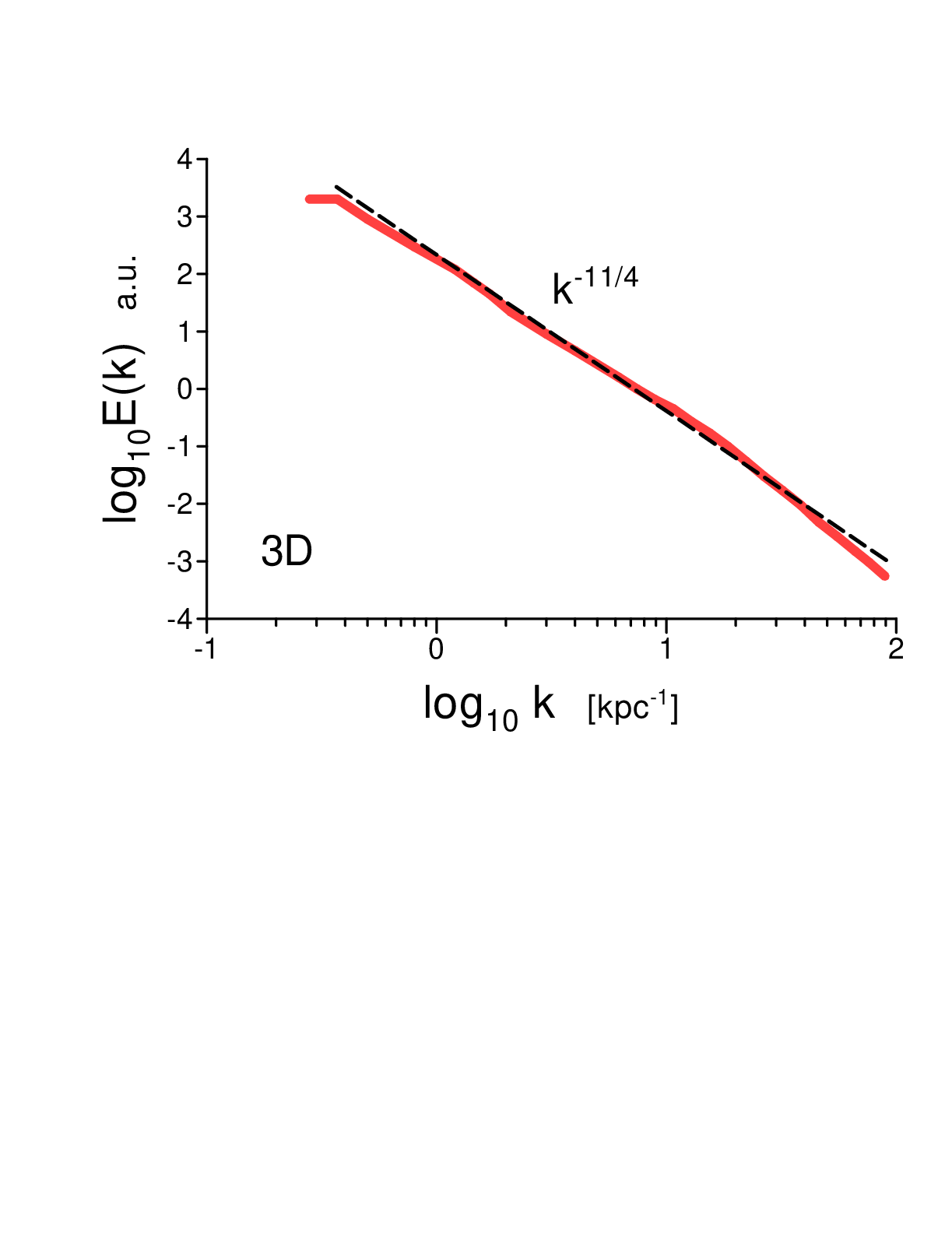} \vspace{-5.3cm}
\caption{The mean power spectrum of the parallel to line-of-sight component of the magnetic field $B_{\parallel}$ for the Milky Way.} 
\end{figure}

   Fig. 8 displays the reconstructed power spectrum of  $B_{\parallel } $ (data from Fig. 4 of Ref.) alongside the reference slope  $-11/4 $. Restricting the inversion to a cube of  $d \leq 1.25\text{ kpc} $ and masking the Galactic plane avoids severe dust extinction and long-path confusion, though it excludes the dense midplane where large-scale dynamo effects peak. Furthermore, because Faraday rotation probes only  $B_{\parallel } $, recovering the full 3D vector field  $\mathbf{B} $ will require joint inversion with 3D dust-polarization tomography tracing the plane-of-sky component  $B_{\perp } $.
Crucially, this spatial masking also resolves an apparent discrepancy between local 3D inversions and full-sky WMAP and Planck foreground analyses, which exhibit stretched-exponential distributed-chaos spectra,  $E(k) \propto \exp\left[-(k/k_\beta)^\beta\right] $ with  $\beta = 1/5 $ and  $\beta = 1/4 $ [36]. As shown in Section IX.C, both spectral forms stem from the same underlying helical amplitude scaling  $b_k \propto k^\alpha $ ( $\alpha = 1/8 $ and  $\alpha = 1/6 $), differing only in whether the cascade parameters ( $\varepsilon  $,  $\varepsilon _{h} $, and  $V_{A} $) remain statistically stationary or fluctuate strongly across the sampled ensemble. In the dense midplane, intense  $\alpha\text{--}\Omega $ dynamo activity, spiral-arm shocks, and overlapping supernova remnants compress and process the plasma (analogous to the post-shock outer heliosheath in Fig. 6) while multi-kiloparsec line-of-sight integration in all-sky maps superposes regions with widely varying local parameters, randomizing the characteristic scale  $k_c $ into a distributed-chaos state. By excluding the dense midplane and resolving the local off-plane volume in 3D, Ref. [43] isolates a quiescent, statistically uniform ensemble of  $\sim 100\text{ pc} $ domains in which the $k^{-3/4} $ scaling projection survives.

\subsection{Galaxy Clusters}

 A recent compilation of 3D magnetic spectral indices derived from Faraday rotation measure (RM) observations across galaxy clusters and groups (Table A.1 of Ref. \cite{lai}) reveals two attracting regimes:  $n_{3\text{D}} \approx -2.75 $ ( $-11/4 $) and  $n_{3\text{D}} \approx -3.6 $. Under the isotropic shell transformation ( $n_{1\text{D}} = n_{3\text{D}} + 2 $), these correspond to 1D spectral indices of  $n_{1\text{D}} \approx -3/4 $ and  $n_{1\text{D}} \approx -1.6 $, respectively.\\
 
 Notably, the three systems locked onto the  $n_{3\text{D}} \approx -2.75 $ attractor (Fornax, Hydra A, and A2199) are relaxed, low-mass, or cool-core systems hosting extended central radio galaxies. These systems have their magnetic fields energized primarily by central active galactic nucleus (AGN) jets and inflated radio lobes (as highlighted for Fornax in Ref. \cite{lai}). These helical, magnetically dominated jets and bubbles inject twisted magnetic flux directly at the upper outer scale of the turbulent cells,  $\Lambda _{\max } $. Because helicity is injected at the largest available turbulent scale, there is no room for an inverse cascade; the plasma is forced to advect magnetic helicity downward to smaller scales ( $k > 2\pi/\Lambda_{\max} $), establishing a decoupled 1D magnetic spectrum  $E_b(k) \propto k^{-3/4} $ ( $n_{3\text{D}} = -11/4 = -2.75 $).
In Hydra A, for example, Laing et al. [45] report a 3D magnetic spectral index of  $n_{3\text{D}} \approx -2.77 \pm 0.10 $ explicitly characterizing the large-scale (low-wavenumber) regime. In their RM structure-function modeling of a high-brightness sub-region of the North lobe located  $5\text{--}15\text{ arcsec} $ ( $\approx 5.3\text{--}16\text{ kpc} $) from the nucleus, the authors parameterize a broken power law, assigning  $n_{3\text{D}} \approx -2.77 $ to the low-wavenumber slope while restricting any steepening to wavenumbers  $k > 5.3\text{ kpc}^{-1} $ (physical scales  $\lambda = 2\pi/k < 1.2\text{ kpc} $). Because their fit spans large angular separations relative to the  $0.3\text{ arcsec} $ ( $\approx 0.32\text{ kpc} $) convolving beam, the resulting power-law index is governed by large-scale, helicity-injected fluctuations. 

  Conversely, massive merging clusters lacking a dominant central helical source (or driven vigorously by non-helical merger shocks and sloshing) undergo the dynamical  first-order transition onto the fast supercritical branch, explaining the steeper  $n_{3\text{D}} \approx -3.6 $ ( $n_{1\text{D}} \approx -1.6 $, intermediate between  $-3/2 $ and  $-5/3 $) attractor.\\
  
 Observing the same  $E_{3\text{D}}(k) \propto k^{-11/4} $ ( $E_{1\text{D}}(k) \propto k^{-3/4} $) signature across three vastly different astrophysical environments (the outer heliosheath, the local Galactic ISM, and relaxed, AGN-magnetized galaxy clusters) alongside the supercritical  $k^{-11/3} $ ( $k^{-5/3} $) branch in non-helical, strongly driven systems points to a universal pair of MHD cascade attractors. Whenever helical magnetic fields are injected at the outer scale of sub-Alfvénic, patchy plasma domains, topological conservation locks the magnetic spectrum onto the  $k^{-3/4} $ ( $n_{3\text{D}} = -11/4 $) branch; conversely, strong non-helical driving triggers the dynamical first-order transition onto the fast Kolmogorov-like branch.

 \section{Summary and Discussion}

   This paper presents a physical framework for the Alfvénic solar wind based on a cellular network of spatial domains, providing a structural alternative to purely statistical MHD models. We posit that the fundamental building block of the solar wind is a discrete plasma domain possessing internal rotational symmetry in the plane perpendicular to its local mean magnetic field. This local symmetry (a remnant of the full 3D isotropy) provides the foundational basis for two things. First, a trigonometric accounting of the angle $\theta_{VB}$ between local mean velocity and magnetic field, using the effective cross-section of colliding counterpropagating Alfv\'en wave packets. Second, a bi-modal treatment of the nonlinear ($\tau_{\perp}=\ell_{\perp}/v_{\ell} =|\sin \theta_{VB}| \ell/v_{\ell} $) and Alfv\'enic ($\tau_{\parallel} = |\cos \theta_{VB}| \ell/V_A$) causal timescales.\\
   
 By applying the dual principles of causality and cascade stability (to the Landau fluctuations of the energy fluxes), we show that the system naturally selects the most efficient cascade rate to prevent intermittent energy flux accumulations. The framework specifically addresses the strong Alfv\'enic imbalance (normalized cross-helicity $\sigma_c \approx 1$) typical of pristine solar-originating domains, where the suppression of nonlinear interactions is regulated by the Alfvénic causal limit. In domains with quasi-parallel sampling, this limit yields a $k^{-2}$ spectrum. Within the magneto-inertial range, a stochastic renewal process with bimodal waiting-time probability renormalizes the cascade time, yielding a two-branch solution: one (slow) branch with a statistically robust anisotropic IK-like $k^{-3/2}$ spectrum and another (fast) branch with a statistically marginal anisotropic Kolmogorov-like spectrum $k^{-5/3}$. \\

   This bimodal cascade structure is not static, however: as the solar wind expands the persistent drive of the expansion triggers a dynamical first-order phase transition in localized, balanced domains ($\sigma_c \approx 0$). These "supercritical" regions switch from the `slow' statistically robust IK-like cascade to a statistically marginal but fast cascade with a Kolmogorov-like $k^{-5/3}$ spectrum to maintain flux stability. These energetic, supercritical $k^{-5/3}$ droplets possess large spatial gradients and progressively (with increasing $R$) dominate the solar wind. \\     
   
   It is shown that the switchbacks can induce heterogeneous nucleation and considerably accelerate the homogeneous Landau/flux-fluctuation phase transition, because they are ready-made supercritical droplets and the system does not need to wait for an unlikely Landau fluctuation when a seed is already sitting there.\\
 
 The difference in the magnetic energy domination over the kinetic energy before and after the phase transition results in the apparent (artificial) decoupling of the velocity and magnetic power spectra obtained from the raw (non-sampling) measurements produced by the probes onboard the spacecraft. These time series actually average (mix) the subcritical and supercritical spatial domains. This results in an apparently faster (at smaller $R$) transition to the global $k^{-5/3}$ spectrum in the magnetic field than in the velocity field.\\
 
   The role of magnetic helicity in the solar wind dynamics is usually underestimated. This can be related to the presence of a strong mean magnetic field (see Introduction). However, the solar wind is rich in magnetic helicity injected by the Sun and by local solar wind dynamics. The magnetic helicity injected by the Sun at the largest scales has no room for the inverse (favorable) cascade and, therefore, is forced by the forward energy cascade downward -- both in the reservoir (largest scales) and in the magneto-inertial range (see, for the latter, Ref. \cite{b1}). It is shown that in the former case, spectra of magnetic and velocity fields are strongly decoupled, with a much flatter magnetic energy spectrum in the reservoir $E(k) \sim k^{-3/4}$ compared to the kinetic energy spectrum $E(k) \sim k^{-3/2}$ (also in the reservoir). For the switchbacks inside the upper solar corona, the large-scale spectra decouple into $k^{-2/3}$ magnetic spectrum and a $k^{-5/3}$ kinetic spectrum. A deep relationship of this approach to the distributed chaos approach to the magneto-inertial range dominated by magnetic helicity \cite{b1} is also discussed. \\
   
   All above-mentioned heliospheric results are supported by comparison with in situ observations from the Voyager, Ulysses, Helios, Wind, and PSP missions across heliocentric distances from $0.056$ to $140~\mathrm{AU}$.\\

Finally, extending this framework beyond the heliosphere shows that the same cascade physics governs astrophysical plasmas on galactic and intracluster scales. In the Milky Way, magnetic helicity conservation combined with large-scale injection by Coriolis-twisted superbubbles drives a two-way helicity cascade -- an inverse cascade sustaining the large-scale Galactic field alongside a direct cascade establishing the $E_{1\mathrm{D}}(k) \propto k^{-3/4}$ ($n_{3\mathrm{D}} \approx -11/4$) magnetic power spectrum observed in 3D Faraday-rotation reconstructions of the local interstellar medium. The same $n_{3\mathrm{D}} \approx -2.75$ scaling emerges in relaxed, cool-core galaxy clusters where central AGN jets inject helicity directly at the outer turbulent scale, whereas non-helical turbulence in massive merging clusters drives the plasma into the mixed Kolmogorov/Iroshnikov--Kraichnan-like regime ($n_{3\mathrm{D}} \approx -3.6$).

 \section{Acknowledgments} 
 
 I thank C. H. K. Chen and F. Loi for consultations.

\end{document}